\documentclass[lettersize,journal]{IEEEtran}
\usepackage{amsmath,amssymb,amsfonts}
\usepackage{algorithmic}
\usepackage{algorithm}
\usepackage{array}
\usepackage[caption=false,font=normalsize,labelfont=sf,textfont=sf]{subfig}
\usepackage{textcomp}
\usepackage{stfloats}
\usepackage{url}
\usepackage{verbatim}
\usepackage{graphicx}
\usepackage{xcolor}
\usepackage{hyperref}

\usepackage{amsthm}
\usepackage{cite}
\usepackage{array, makecell}
\usepackage{mathtools}
\newtheorem{theorem}{Theorem}   
\newtheorem{lemma}{Lemma}

\newtheorem{definition}{Definition}

\newtheorem{assumption}{Assumption}

\usepackage[font=small,labelfont=bf]{caption}
\begin{document}




\title{FLAME: A Federated Learning Approach for Multi-Modal RF Fingerprinting}

\author{Kasra Borazjani$^{*}$, Kiarash Kianfar$^{*}$, Seyyedali Hosseinalipour, and Rajeev Sahay

\thanks{$^{*}$K. Borazjani and K. Kianfar contributed equally to this work.}
\thanks{Code is publicly available at \url{https://github.com/KasraBorazjani/flame}.}
\thanks{K. Borazjani and S. Hosseinalipour are with the Department of Electrical Engineering, State University of New York at Buffalo, Buffalo, NY, 14260 USA. E-mail: \{kasrabor,alipour\}@buffalo.edu.}
\thanks{K. Kianfar and R. Sahay are with the Department of Electrical and Computer Engineering, University of California San Diego, San Diego, CA, 92093 USA. E-mail: \{kkianfar,r2sahay\}@ucsd.edu.}
\thanks{This work was supported in part by the UC San Diego Academic Senate under grant RG114404 and in part by the National Science Foundation (NSF) under grant ECCS-2512912, ECCS-2512911, and SaTC-2513164.}
\thanks{Copyright (c) 20xx IEEE. Personal use of this material is permitted. However, permission to use this material for any other purposes must be obtained from the IEEE by sending a request to pubs-permissions@ieee.org.}}



\maketitle

\begin{abstract} Authorization systems are increasingly relying on processing radio frequency (RF) waveforms at receivers to fingerprint (i.e., determine the identity of) the corresponding transmitter. Federated learning (FL) has emerged as a popular paradigm to perform RF fingerprinting in networks with multiple access points (APs), as they allow effective deep learning-based device identification without requiring the centralization of locally collected RF signals stored at multiple APs. Yet, FL algorithms that operate merely on in-phase and quadrature (I/Q) time samples incur high convergence rates, resulting in excessive training rounds and inefficient training times. In this work, we propose FLAME: an FL approach for multi-modal RF fingerprinting. Our framework consists of simultaneously representing received RF waveforms in multiple complementary modalities beyond I/Q samples in an effort to reduce training times. We theoretically demonstrate the feasibility and efficiency of our methodology and derive a convergence bound that incurs lower loss and thus higher accuracies in the same training round in comparison to single-modal FL-based RF fingerprinting. Extensive empirical evaluations validate our theoretical results and demonstrate the superiority of FLAME in comparison to multiple considered baselines. 
\end{abstract}


\section{Introduction}

\IEEEPARstart{A}{ll} pieces of transmitting hardware inherently contain slight physical differences due to imperfections in their manufacturing processes. Due to these hardware imperfections, radio transmitters have slight variations in their transmitted radio frequency (RF) waveforms, reflected in the channel state information (CSI) of the received signal \cite{PARADIS_non_dl}. RF fingerprinting is a technique that utilizes these unchanging variations in transmitted signals to identify the transmitting device. Such RF fingerprinting techniques are vital in both military and civilian applications, where accurate device identification is required to prevent receivers from servicing unauthorized or adversarial transmitters \cite{adv_fl}. 


Current state-of-the-art methods for RF fingerprinting are largely comprised of deep learning approaches \cite{dl_fingerprinting1,dl_fingerprinting2,dl_fingerprinting3,dl_fingerprinting4,dl_fingerprinting5}, which significantly outperform statistical signal processing methods such as maximum likelihood classifiers \cite{ml1,ml2,ml3}. However, deep learning approaches are difficult to scale in next-generation communications, where radio waveforms are typically stored at multiple access points (APs) within a wireless network and require the aggregation of gigabytes to terabytes of data to a centralized location before model training can occur. Not only would such data aggregation be computationally costly, but it also increases the potential of data leakage during the aggregation process. 

Federated learning (FL), a distributed and privacy-preserving deep learning paradigm, has recently emerged as a potential solution for RF fingerprinting to mitigate the challenges of centralized deep learning approaches \cite{fl_fingerprinting,fl_fingerprinting2,fl_fingerprinting3}. The FL RF fingerprinting approach trains local models at each AP on locally stored RF data at each AP. Each AP, after performing local training, then transmits only the updated model parameters (which are often much smaller than the size of the fingerprinting data) to a centralized location (e.g., a coordinating server) for aggregation. The central server then aggregates its received model parameters into a global model and transmits the updated parameters of the global model to each AP to resume the training process. In this fashion, the locally collected fingerprinting data at each AP never leaves its stored location, reducing communication overhead and improving data security. 

Despite its benefits, FL poses two key challenges in large-scale RF fingerprinting networks. First, RF fingerprinting data across APs are often heterogeneous and, thus, each AP only contains fingerprinting data corresponding to a subset of transmitters attempting to be learned by the global model. As a result, the global model struggles to generalize to all transmitters, impeding overall performance. Second, the central server in FL requires frequent aggregation from each AP, incurring high communication overhead. Together, these challenges require performing (i) a low number of model parameter updates during each training round to prevent the weights of each local model from diverging too significantly from the global model \cite{non_iid} and (ii) a large number of FL training iterations \cite {comm_ch} in order to achieve high RF fingerprinting performance.
Hence, traditional FL algorithms are slower to learn new devices introduced in non-independent and identically distributed (non-IID) settings~\cite{FL_nonIID_lowerPeformance}, and in distinguishing between trusted and untrusted devices. While prior work has explored multi-modal representations for RF fingerprinting in centralized learning settings~\cite{mm_amc1,mm_fingerprinting_dl}, extending such approaches to FL is non-trivial due to challenges unique to FL, including data heterogeneity, constrained local data, and limited cross-client coordination. Moreover, existing multi-modal FL frameworks have largely focused on non-RF data types~\cite{mm_FL_recent,MM_FL_IoT,MM_FL_ImageBased,FL_MM_fall_detection,MM_FL_concept}, leaving multi-modal RF fingerprinting in federated environments underexplored. These gaps motivate our proposed FLAME framework, which is specifically designed to integrate complementary RF modalities within an FL paradigm.

In this work, we develop an FL approach for multi-modal RF fingerprinting (FLAME). FLAME consists of using complementary signal information at each AP, in the form of multiple signal modalities, to perform local training prior to global aggregation. We show that, under our proposed framework, the variance of the global loss is proportionally reduced by a factor consistent with the number of modalities used at each AP. As a result, a larger number of local parameter updates can occur at each AP in FLAME, since the variance of the aggregated loss is reduced in comparison to using a single modality. In addition, the FLAME training process requires fewer communication rounds between APs and the central server since the variance of the loss at each AP is reduced, thus increasing the tolerance with which the weights of the local models can diverge prior to global aggregation. We theoretically and empirically show the benefits of FLAME in terms of efficiency and fingerprinting performance in comparison to previously proposed FL approaches for RF fingerprinting.

\textbf{Summary of Contributions:} Our contributions, in comparison to related work (discussed in Sec. II), can be summarized as follows:
\begin{enumerate} 
    \item \textbf{Development of FLAME} (Sec. III-A -- Sec. III-C): We develop FLAME, an efficient federated learning framework for multi-modal RF fingerprinting. \emph{To the best of our knowledge, this is the first work that jointly integrates multi-modal RF signal representations within a federated learning paradigm for RF fingerprinting.} Specifically, FLAME leverages complementary RF representations, including the raw I/Q waveform, frequency-domain components obtained through the discrete Fourier transform (DFT), and amplitude–phase signal representations, enabling improved device discrimination while preserving data privacy in distributed environments.
    
    \item \textbf{Theoretical Feasibility of FLAME} (Sec. III-D): We theoretically analyze the feasibility of FLAME. Specifically, we show that incorporating multiple complementary RF signal modalities improves the convergence  and leads to improved fingerprinting performance compared to existing single-modal FL-based RF fingerprinting approaches.
    
    \item \textbf{Empirical Evaluation} (Sec. IV-A -- IV-J): We conduct extensive experiments on multiple real-world RF fingerprinting datasets to validate our theoretical findings. The results demonstrate that FLAME achieves higher performance compared to its single-modal FL counterparts. Furthermore, we show that the performance improvements introduced by the multi-modal framework remain consistent across multiple federated optimization algorithms, including FedAvg, FedProx, and SCAFFOLD.
\end{enumerate}


\begin{figure*}[t!]
    \centering
    \includegraphics[width=2.0\columnwidth]{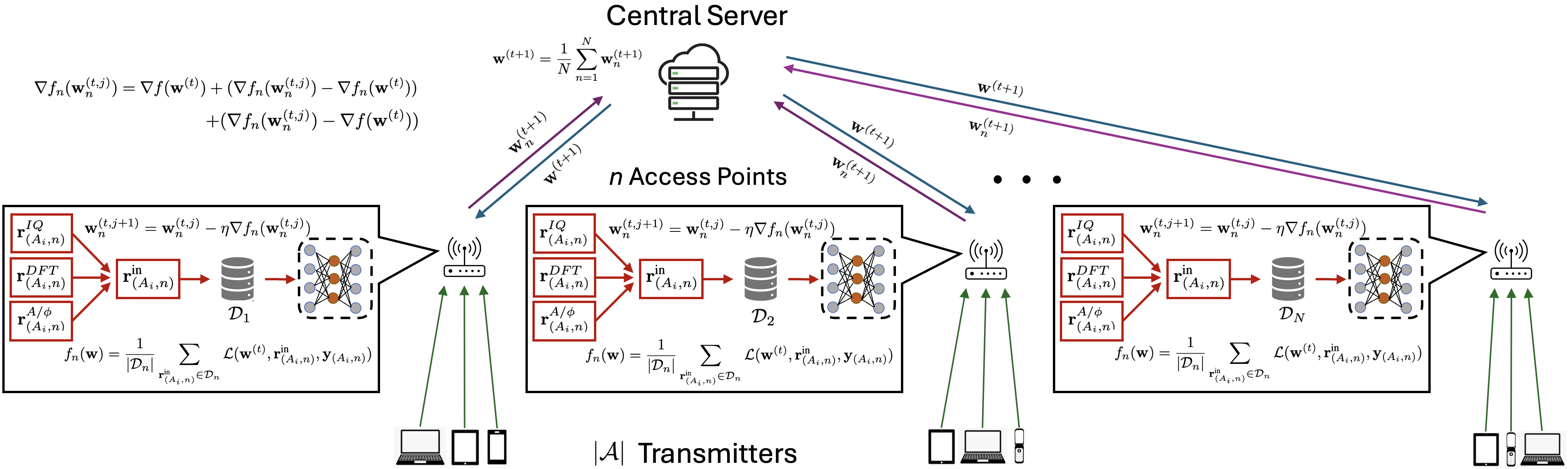}
    \caption{The FLAME system diagram consisting of $N$ access points (APs) servicing $|\mathcal{A}|$ transmitters.}
    \label{sys_diagram}
\end{figure*}




\section{Related Works} \label{related_work}






RF fingerprinting primarily relies on performing device authentication based on the transient response of RF signals \cite{non_dl_transient_extraction}. Early classification methods consisted of correlation detection \cite{preamble_based_detection_non_dl}, signal strength measurements from multiple access points \cite{non_dl_tracking_signal_strength}, and maximum likelihood approaches \cite{ml1,ml2,ml3}. However, such methods struggle to yield high accuracy for effective device classification. Recent data-driven machine learning approaches, on the other hand, have been successful in achieving high RF fingerprinting performance directly on the transient response of received signals \cite{dl_fingerprinting1,dl_fingerprinting2, dl_fingerprinting3, dl_fingerprinting4, dl_fingerprinting5, RFID_authentication,Semi-supervised_learning}. Yet, these methods primarily rely on the traditional, centralized training approach, which has potential limitations. Specifically, the aggregation of RF data from multiple wireless access points is not feasible for large local datasets and increases chances of data leakage, raising privacy and security concerns.

Federated learning (FL) has emerged as a promising distributed RF fingerprinting approach, allowing for model training at local access points directly, and thus eliminating the need for data centralization. Multiple frameworks for FL have been explored for RF fingerprinting \cite{fl_fingerprinting,fl_fingerprinting2,fl_fingerprinting3} as well as long-range (LoRa) communications \cite{LoRa_FL_RF}. These studies have shown the improvements of FL in terms of data privacy. However, FL has shown lower performance than centralized training due to the heterogeneity of data distributions at various access points as well as the reliance on single-modal (i.e., I/Q-based) processing as opposed to multi-modal-based FL as we propose. 

In addition to FL applications in wireless communications, several general FL algorithms have been proposed to mitigate performance degradation caused by data heterogeneity. The widely used FedAvg algorithm~\cite{FedAvg} aggregates clients' local model updates via weighted averaging but often struggles under non-IID data settings due to client drift. FedProx~\cite{fL_heterogeneous} extends FedAvg by introducing a proximal regularization term in the local objective to stabilize training under non-IID data across clients, while SCAFFOLD~\cite{scaffold} addresses client drift through controlled aggregation using control variates. More recent approaches, such as the normalized update strategy in~\cite{wang2020tacklingobjectiveinconsistencyproblem}, further improve convergence stability by accounting for differences in local training dynamics and client participation. Additionally, FedAMP~\cite{huang2021personalized} enables personalized FL through attentive message passing, adaptively facilitating collaboration among clients with similar model parameters under non-IID data distributions.
While these methods address statistical data heterogeneity, they are primarily designed for unimodal learning and do not explicitly model modality-specific feature extraction. In contrast, FLAME employs modality-specific encoders and a late fusion strategy to integrate complementary signal representations, enabling more effective learning under multi-modal RF data heterogeneity.

Meanwhile, separate studies have worked with multi-modal RF data, showing the performance improvements obtained by utilizing multiple modalities of the same data for centralized deep learning, primarily for automatic modulation classification (AMC) \cite{mm_amc1,mm_amc2,mm_amc3}, but also for RF fingerprinting \cite{mm_fingerprinting_dl}. Multi-modal-based FL approaches have also emerged in various applications outside of wireless communications, improving FL performance, compared to single-modal training, in a variety of settings \cite{MM_FL_concept,MM_FL_IoT,FL_MM_fall_detection}. However, multi-modal FL, encompassing its unique challenges such as data heterogeneity across access points, has not been explored for RF fingerprinting. Thus, in this work, we propose the first multi-modal-based FL framework for RF fingerprinting and demonstrate its efficacy both theoretically and empirically.




\section{Methodology} \label{methods}

In the following, we state our signal model (Sec. \ref{sig_mod}) and describe our multi-modal (Sec. \ref{mm_rep}) and  FL  framework (Sec. \ref{clf_mod}). We then demonstrate the feasibility of our methodology through a convergence analysis (Sec. \ref{conv_an}). An overview of our FLAME methodology is shown in Fig. \ref{sys_diagram}. 

\subsection{Signal Modeling} \label{sig_mod}


We consider an FL framework consisting of $\mathcal{A} = \{A_{1}, A_{2}, \ldots, A_{|\mathcal{A}|}\}$ transmitters and $n = 1, 2, \ldots, N$ Access Points (APs), where each AP contains a local dataset denoted by $\mathcal{D}_{n}$ consisting of $|\mathcal{D}_{n}|$ samples. At each AP, $\mathcal{D}_{n}$ is comprised of a set of signals, received from a subset of transmitters $\mathcal{A}_{n} = \{A_{1}, A_{2}, \ldots, A_{|\mathcal{A}_{n}|}\}$, where $\mathcal{A}_{n} \subset \mathcal{A}$ and, thus, $|\mathcal{A}_{n}| < |\mathcal{A}|$. We further assume that each transmitter is only serviced by a single and unique AP and, therefore, $\mathcal{D}_{1} \cap \mathcal{D}_{2} \cap \ldots \cap \mathcal{D}_{N} = \emptyset$. Here, we consider two distinct scenarios: (i) $\mathcal{A}_{1} \cap \mathcal{A}_{2} \cap \ldots \cap \mathcal{A}_{N} = \emptyset$ and (ii) $\mathcal{A}_{1} \cap \mathcal{A}_{2} \cap \ldots \cap \mathcal{A}_{N} \neq \emptyset$. Practically, (i) corresponds to non-independent and identically distributed (non-IID) transmitters, where each AP only receives transmissions from a specific subset of transmitters and is thus only exposed to a subset of transmitter fingerprints, and (ii) corresponds to an IID setting, where a transmitter may transmit to different APs and thus multiple APs contain fingerprints from multiple transmitters.    

Formally, the received signal from the $i^{\text{th}}$ transmitter, $A_{i}$, at AP $n$ is modeled by
\begin{equation}
    \mathbf{r}_{(A_{i}, n)} = h_{A_{i}}(\mathbf{s}, t) + \mathbf{n}, 
\end{equation}
where $\mathbf{r}_{(A_{i}, n)} = [r_{(A_{i}, n)}[0], \cdots, r_{(A_{i}, n)}[\ell -1]]^{\text{T}} \in \mathbb{C}^{\ell}$, $\ell$ is the length of the received signal's observation window, $\mathbf{n} \in \mathbb{C}^{\ell}$ represents complex additive white Gaussian noise (AWGN), $\mathbf{s}$ is the set of transmitted symbols and $h_{A_{i}}$ represents the time-variant RF fingerprint (of $A_{i}$), which captures the transmitter hardware fingerprints and wireless channel effects. We assume that the channel distribution between each transmitter and its receiving AP is IID. The FL objective is to learn a global RF fingerprinting classifier to predict $A_{i} \in \mathcal{A}$, given $\mathbf{r}_{(A_{i}, n)}$, by training all local models at each AP.

The selected modalities used in model training should capture complementary characteristics of the transmitted signal. Specifically, prior centralized deep learning approaches for RF fingerprinting have shown that distinct signal properties are emphasized across different representations, such as amplitude, phase, and spectral characteristics~\cite{multiple_format}. In this work, we extend this principle to the federated learning setting, where leveraging a richer feature space can help mitigate the adverse effects of data heterogeneity and accelerate convergence~\cite{MM_FL_IoT, MM_FL_concept}. Unlike prior centralized multi-modal architectures that rely on early fusion or deep backbones for each modality's feature extraction~\cite{mm_amc1,mm_fingerprinting_dl}, FLAME adopts a lightweight, modality-specific encoding strategy with late fusion, enabling more robust and efficient learning under federated constraints. Specifically, we utilize I/Q, frequency-domain, and phasor representations extracted from the same received signal to jointly capture complementary information.

\begin{figure*}[t]
    \centering
    \includegraphics[width=2.0\columnwidth]{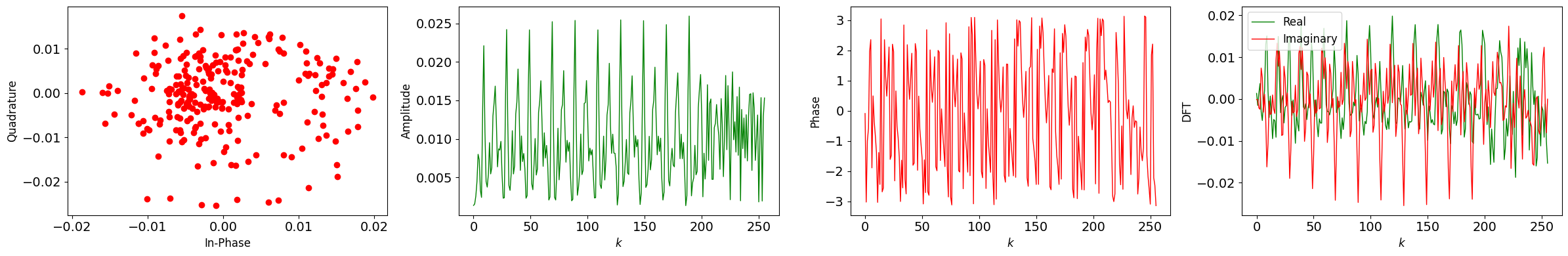}
    \caption{A visualization of each considered modality, which shows Modality 1 (first), Modality 2 (second), and Modality 3 (third and fourth).}
    \label{sigs}
\end{figure*}



\subsection{Multi-Modal Domain Representations} \label{mm_rep}

We represent each received signal, $\mathbf{r}_{(A_{i}, n)}$, in three different modalities in order to model each signal using discriminative features for accurate RF fingerprinting. Specifically, we consider the (i) in-phase and quadrature (I/Q) time-domain representation of the received signal at complex baseband, (ii) frequency-domain representation obtained via the Discrete Fourier Transform (DFT) of the received complex baseband signal, and (iii) amplitude and phase of the received complex baseband signal. Each representation has been independently used to perform classification in various wireless settings, including, but not limited to RF fingerprinting \cite{mm_amc3,multiple_format}, but no work has jointly used all three considered modalities for RF fingerprinting. We detail our methodology used to arrive at each of the specific modalities below. 

\noindent \textbf{(Modality 1)} We first model each received signal, $\mathbf{r}_{(A_{i}, n)} \in \mathbb{C}^{\ell}$, in its baseband form. Following prior work \cite{ota_amc,fl_fingerprinting2}, we map each baseband I/Q signal to a two-dimensional real matrix, $\mathbf{r}_{(A_{i}, n)} \in \mathbb{C}^{\ell} \rightarrow \mathbf{r}_{(A_{i}, n)}^{\text{I/Q}} \in \mathbb{R}^{\ell \times 2}$ where the first and second column of $\mathbf{r}_{(A_{i}, n)}^{\text{I/Q}}$ represent the real and imaginary components, respectively, of $\mathbf{r}_{(A_{i}, n)}$ so that they can be used with real-valued neural networks. 

\noindent \textbf{(Modality 2)} Next, we model $\mathbf{r}_{(A_{i}, n)} = [r_{t}[0],\ldots,r_{t}[\ell - 1]]^{T}$ using the frequency components obtained from its discrete Fourier transform (DFT). Specifically, the $p^{\text{th}}$ frequency component of the DFT of $\mathbf{r}_{(A_{i}, n)}$ is given by 
\begin{equation} \label{dft}
    \mathbf{R}_{(A_{i}, n)}[p] = \sum_{k=0}^{\ell - 1} \mathbf{r}_{(A_{i}, n)}[k] e^{-\frac{j2\pi}{\ell}pk}, \hspace{2mm} p = 0,\ldots, \ell - 1, 
\end{equation}
\noindent where $\mathbf{R}_{(A_{i}, n)} = [\mathbf{R}_{(A_{i}, n)}[0],\ldots,\mathbf{R}_{(A_{i}, n)}[\ell - 1]]^T \in \mathbb{C}^{\ell}$ contains all frequency components of $\mathbf{r}_{(A_{i}, n)}$. Similar to Modality 1, we map $\mathbf{R}_{(A_{i}, n)}$ to a two-dimensional real matrix, $\mathbf{R}_{(A_{i}, n)} \in \mathbb{C}^{\ell} \rightarrow \mathbf{r}_{(A_{i}, n)}^{\text{DFT}} \in \mathbb{R}^{\ell \times 2}$ where the first and second column of $\mathbf{r}_{(A_{i}, n)}^{\text{DFT}}$ represent the real and imaginary components, respectively, of $\mathbf{R}_{(A_{i}, n)}$. 


\noindent  \textbf{(Modality 3)} Finally, we model $\mathbf{r}_{(A_{i}, n)}$ using its amplitude and phase for each time sample. Specifically, we determine 
\begin{equation} \label{amp}
    r_{(A_{i}, n)}^{\text{A}}[k] = \sqrt{(\Re(r_{(A_{i}, n)}[k]))^2 + (\Im(r_{(A_{i}, n)}[k]))^2}
\end{equation}
and 
\begin{equation} \label{phase}
    r_{(A_{i}, n)}^{\phi}[k] = \text{tan}^{-1}\bigg{(}\frac{\Im(r_{(A_{i}, n)}[k])} {\Re(r_{(A_{i}, n)}[k])}\bigg{)}
\end{equation}
for $k = 0,\ldots, \ell - 1$, where $\Re(\cdot)$ and $\Im(\cdot)$ represent the real and imaginary component of $(\cdot)$, respectively. For each signal, $\mathbf{r}_{(A_{i}, n)}$, we store the result of (\ref{amp}) and (\ref{phase}) in a real-valued two-dimensional matrix denoted by $\mathbf{r}_{(A_{i}, n)}^{\text{A}/\phi} \in \mathbb{R}^{\ell \times 2}$, where the first and second column of $\mathbf{r}_{(A_{i}, n)}^{\text{A}/\phi}$ represent the amplitude and phase, respectively, of $\mathbf{r}_{(A_{i}, n)}$.

Fig. \ref{sigs} visualizes examples of $\mathbf{r}_{(A_{i}, n)}^{\text{I/Q}}$, $\mathbf{r}_{(A_{i}, n)}^{\text{DFT}}$, and $\mathbf{r}_{(A_{i}, n)}^{\text{A}/\phi}$ for signals received with various RF fingerprints. For each local dataset at AP $n$, we compute an $M$ channel input for each signal, denoted by $\mathbf{r}_{(A_{i}, n)}^{\text{in}} \in \mathbb{R}^{\ell \times 2 \times M}$, where $M$ denotes the number of modalities, each consisting of its own encoder (e.g., jointly using Modality 1, 2, and 3 would result in $M = 3$ since three modalities are used). 

\subsection{Classifier Modeling} \label{clf_mod}


We denote a deep learning classifier as $z_{\mathbf{w}}: \mathbf{r}_{(A_{i}, n)}^{\text{in}} \rightarrow \hat{\mathbf{y}}_{(A_{i}, n)}$, where the classifier is parametrized by $\mathbf{w}$ and trained to fingerprint the input, $\mathbf{r}_{(A_{i}, n)}^{\text{in}} \in \mathbb{R}^{\ell \times 2 \times M}$, to $\hat{\mathbf{y}}_{(A_{i}, n)} \in \mathbb{R}^{|\mathcal{A}|}$, which denotes the predicted transmitter. Each AP shares the same deep learning architecture. The objective of the FL network is to find a model parameter $\mathbf{w}$ that minimizes the loss between $\hat{\mathbf{y}}_{(A_{i}, n)}$ and the true transmitter $\mathbf{y}_{(A_{i}, n)}$. Formally, this can be expressed as
\begin{equation} \label{fl_obj}
     \underset{\mathbf{w}}{\text{min}} \hspace{0.25cm} f(\mathbf{w}), \hspace{0.25cm} \text{where} \hspace{0.25cm} f(\mathbf{w}) \coloneqq \frac{1}{N}\sum_{n=1}^{N} f_{n}(\mathbf{w}), 
\end{equation}
\begin{equation}\label{eq:LocLoss}
    f_{n}(\mathbf{w}) =  \frac{1}{|\mathcal{D}_{n}|} \sum_{\mathbf{r}^{\text{in}}_{(A_{i}, n)} \in \mathcal{D}_{n}} \mathcal{L}(\mathbf{w}, \mathbf{r}^{\text{in}}_{(A_{i}, n)}, \mathbf{y}_{(A_{i}, n)})
\end{equation}
and $\mathcal{L}(\mathbf{w}, \mathbf{r}^{\text{in}}_{(A_{i}, n)}, \mathbf{y}_{(A_{i}, n)})$ is the local loss function at AP $n$ over $\mathcal{D}_{n}$. 

Since the fingerprinting data is distributed among $N$ access points, (\ref{fl_obj}) is achieved through multiple rounds of local learning, parameter aggregation, and global model synchronization. Specifically, at the beginning of each training round, $t$, the global model transmits its parameters, $\mathbf{w}^{(t)}$, to each AP. AP $n$ then initializes $\mathbf{w}^{(t, 0)}_{n} = \mathbf{w}^{(t)}$ and trains its local model by minimizing its local loss using $\mathcal{D}_{n}$, in which each input signal is constructed by concatenating $M$ modalities, using $J$ stochastic gradient descent (SGD) steps according to 
\begin{equation} \label{local_update}
    \mathbf{w}^{(t, j+1)}_{n} = \mathbf{w}^{(t, j)}_{n} - \eta \widetilde{\nabla} f_{n}(\mathbf{w}^{(t, j)}_{n}),
\end{equation}
where $j = 0, \cdots, J-1$, $\widetilde{\nabla}$ represents the stochastic gradient with respect to $\mathbf{w}^{(t, j)}_{n}$ computed over a mini-batch of datapoints randomly selected from the local dataset of AP, and $\eta$ is the learning rate (i.e., step size). 
Since the local model at AP $n$ is trained on instances of $\mathbf{r}^{\text{in}}_{(A_{i}, n)}$, which contain $M$ modalities, $\mathbf{w}^{(t+1)}_{n}$ accounts for the updated loss from $M$ modalities simultaneously. 



At the termination of the training round, each local device sets $\mathbf{w}^{(t+1)}_{n} = \mathbf{w}^{(t, J)}_{n}$ and returns $\mathbf{w}^{(t+1)}_{n}$, which are the model parameters of AP $n$ after the completion of training round $t$ on $\mathcal{D}_{n}$, to the global server. The global server then performs parameter aggregation by calculating the parameters for the next training iteration according to
\begin{equation} \label{agg_eqn}
    \mathbf{w}^{(t+1)} = \frac{1}{N} \sum_{n=1}^{N} \mathbf{w}^{(t+1)}_{n}. 
\end{equation}
After aggregation, the global server transmits the updated model parameters, $\mathbf{w}^{(t+1)}$, for the next round of training to each AP. This process continues for $T$ training iterations. Our complete FLAME methodology for a single training iteration of RF fingerprinting is detailed in Algorithm \ref{flame}. 


\begin{algorithm}[t] 
   \caption{FLAME methodology at training iteration $t>0$}
   \label{flame}
   \begin{algorithmic}[1] 
        \STATE \textbf{input:} $\mathbf{w}^{(t)}$: global parameters at training iteration $t$ \\ 
        \hspace{9mm} $B$: batch size \\
        \hspace{9mm} $J$: SGD iterations \\
        \hspace{9mm} $\eta$: learning rate at each AP  \\
        
        \FOR{$n = 1, \ldots, N$ (in parallel)}

            \FOR{each $\mathbf{r}_{(A_{i}, n)} \in \mathcal{D}_{n}$}
            
                \STATE Compute $\mathbf{r}_{(A_{i}, n)}^{\text{I/Q}}$, $\mathbf{r}_{(A_{i}, n)}^{\text{DFT}}$, $\mathbf{r}_{(A_{i}, n)}^{\text{A}/\phi} \in \mathbb{R}^{\ell \times 2}$
                
                \STATE Construct $\mathbf{r}_{(A_{i}, n)}^{\text{in}} \in \mathbb{R}^{\ell \times 2 \times M}$ where $M = 3$
                
                \STATE $\mathcal{D}_{n} = \mathcal{D}_{n} \backslash \{\mathbf{r}_{(A_{i}, n)} \}$ 

                \STATE $\mathcal{D}_{n} = \mathcal{D}_{n} \cup \{\mathbf{r}_{(A_{i}, n)}^{\text{in}}\}$  

            \ENDFOR 

            \STATE $\mathbf{w}^{(t, 0)}_{n} \gets \mathbf{w}^{(t)}$
                
            \FOR{$j = 0, 1, \cdots, J-1$}

                \STATE Form mini-batch of $B$ random samples $\mathcal{B}_{n}^{(j)} \subseteq \mathcal{D}_{n}$

                \STATE {\scriptsize $\widetilde{\nabla} f_{n}(\mathbf{w}^{(t, j)}_{n}) {=}  \frac{1}{|\mathcal{B}_{n}^{(j)}|}\sum_{\mathbf{r}^{\text{in}}_{(A_{i}, n)} \in \mathcal{B}_{n}^{(j)}} \nabla \mathcal{L}(\mathbf{w}^{(t, j)}_{n}, \mathbf{r}^{\text{in}}_{(A_{i}, n)}, \mathbf{y}_{(A_{i}, n)})$}

                \STATE $\mathbf{w}^{(t, j+1)}_{n} \gets \mathbf{w}^{(t, j)}_{n} - \eta \widetilde{\nabla} f_{n}(\mathbf{w}^{(t, j)}_{n})$

            \ENDFOR 

            \STATE $\mathbf{w}^{(t+1)}_{n} \gets \mathbf{w}^{(t, J)}_{n}$
            
        \ENDFOR 

        \STATE $\mathbf{w}^{(t+1)} = \frac{1}{N}\sum_{n=1}^{N}  \mathbf{w}^{(t+1)}_{n}$ 
        
        \RETURN $\mathbf{w}^{(t+1)}$
  
  \end{algorithmic}

\end{algorithm}

\subsection{Convergence Analysis} \label{conv_an}


To perform classification over the introduced multi-modal data, each AP is
assumed to employ a multi-modal architecture composed of \textit{modality-specific
encoders}, a \textit{late-fusion module} (deployed before the decoder), and \textit{a shared decoder}. Subsequently, the local objective of
AP $n$ from ~\eqref{eq:LocLoss} can be equivalently rewritten in terms of the output of the local model as 
\begin{equation}
\hspace{-3mm}
\resizebox{0.92\linewidth}{!}{$
\begin{aligned}
f_n(\mathbf{w})
{=}
\frac{1}{|\mathcal{D}_{n}|} 
\sum_{\mathbf{r}^{\text{in}}_{(A_{i}, n)} \in \mathcal{D}_{n}}
\mathcal{L}
\Big(\mathbf{w},
D_{\boldsymbol{\theta}_d}
\big(
\Phi_{\boldsymbol{\theta}_f}
(
E_{1,\boldsymbol{\theta}_1}(\mathbf{r}_{(A_{i}, n)}^{1}),
\ldots,
E_{M,\boldsymbol{\theta}_M}(\mathbf{r}_{(A_{i}, n)}^{M})
)
\big),
\mathbf{y}_{(A_{i}, n)}
\Big),
\end{aligned}
$}
\label{eq:mm_local_objective}
\hspace{-3mm}
\end{equation}
where $\mathbf{r}_{(A_{i}, n)}^{m}$ is the input associated with modality $m$\footnote{We consider $m\in\{1,2,3\}$ to be the index of the modality, where the indexing is order-invariant (e.g., $m=1$ may correspond to I/Q, $m=2$ may correspond to DFT, and $m=3$ may correspond to $A/\phi$).},
$E_{m,\boldsymbol{\theta}_m}(\cdot)$ is the encoder of modality $m$ parametrized by $\boldsymbol{\theta}_m$,
$\Phi_{\boldsymbol{\theta}_f}(\cdot)$ is the late-fusion module, and
$D_{\boldsymbol{\theta}_d}(\cdot)$ is the shared decoder parameterized by $\boldsymbol{\theta}_d$. The full model
parameter vector is therefore given by
\begin{equation}
\mathbf{w}
=
\left[
\boldsymbol{\theta}_1,\ldots,\boldsymbol{\theta}_M,
\boldsymbol{\theta}_f,
\boldsymbol{\theta}_d
\right].
\label{eq:mm_parameter_vector}
\end{equation}
We thus can write the full local gradient in block form as
\begin{equation}
\hspace{-2mm}
\resizebox{0.92\linewidth}{!}{$
\nabla f_n(\mathbf{w})
=
\left[
\nabla_{\boldsymbol{\theta}_1} f_n(\mathbf{w}),
\ldots,
\nabla_{\boldsymbol{\theta}_M} f_n(\mathbf{w}),
\nabla_{\boldsymbol{\theta}_f} f_n(\mathbf{w}),
\nabla_{\boldsymbol{\theta}_d} f_n(\mathbf{w})
\right].
\label{eq:mm_full_gradient_block}
$}\hspace{-1mm}
\end{equation}
This representation shows that the multi-modal gradient consists of
modality-specific encoder blocks and shared fusion-decoder blocks. Using this form, we provide the following definition to expose the role of each modality in the convergence analysis.

\begin{definition}[Modality-Conditioned Gradient]
Let
$f_{n,m}(\mathbf{w})$ denote the loss induced by modality $m$ at AP $n$. This
loss can be obtained by activating only modality $m$, or equivalently by
measuring the contribution of modality $m$ through masking or ablation in the
fused multi-modal model. We define the corresponding modality-conditioned
gradient as
\begin{equation}
\mathbf{g}_{n,m}(\mathbf{w})
\triangleq
\nabla_{\mathbf{w}} f_{n,m}(\mathbf{w})
\in \mathbb{R}^{d},
\quad d=\dim(\mathbf{w}).
\label{eq:mm_modality_conditioned_gradient}
\end{equation}
Since $\mathbf{g}_{n,m}(\mathbf{w})$ is taken with respect to the full model
parameter vector $\mathbf{w}$, all modality-conditioned gradients have the same
dimension $d=\dim(\mathbf{w})$. Also, in block form,
\begin{equation}
\hspace{-2mm}
\resizebox{0.92\linewidth}{!}{$
\mathbf{g}_{n,m}(\mathbf{w})
=
\left[
\mathbf{0},
\ldots,
\nabla_{\boldsymbol{\theta}_m} f_{n,m}(\mathbf{w}),
\ldots,
\mathbf{0},
\nabla_{\boldsymbol{\theta}_f} f_{n,m}(\mathbf{w}),
\nabla_{\boldsymbol{\theta}_d} f_{n,m}(\mathbf{w})
\right],
$}\hspace{-2mm}
\label{eq:mm_modality_gradient_block}
\end{equation}
where the zero blocks correspond to modality-specific encoders that are not
activated by modality $m$.
\end{definition}

 The
 definition in~\eqref{eq:mm_modality_conditioned_gradient}
embeds each modality-conditioned gradient into the same parameter space
$\mathbb{R}^{d}$. We note that since encoder blocks are disjoint across
modalities, the most informative \textit{cross-modal interactions} occur on the shared
fusion-decoder parameters of the model. Therefore, we introduce
\begin{equation}
\mathbf{g}_{n,m}^{\mathrm{Sh}}(\mathbf{w})
\triangleq
\left[
\nabla_{\boldsymbol{\theta}_f} f_{n,m}(\mathbf{w}),
\nabla_{\boldsymbol{\theta}_d} f_{n,m}(\mathbf{w})
\right],
\label{eq:mm_shared_modality_gradient}
\end{equation}
which denotes the modality-induced gradient on the shared part of the model. Using this introduced quantity, we next provide another definition that captures the cross-modal interaction between the utilized modalities in the model.

\begin{definition}[Local Cross-Modal Interaction]
The local cross-modal interaction coefficient at AP $n$ is defined as
\begin{equation}
\varrho_n(\mathbf{w})
\triangleq
\frac{
2 \sum_{m=1}^{M-1} \sum_{m'=m+1}^{M}
\left\langle
\mathbf{g}_{n,m}^{\mathrm{Sh}}(\mathbf{w}),
\mathbf{g}_{n,m'}^{\mathrm{Sh}}(\mathbf{w})
\right\rangle
}{
(M-1)
\sum_{m=1}^{M}
\left\|
\mathbf{g}_{n,m}^{\mathrm{Sh}}(\mathbf{w})
\right\|^2
+\epsilon
},
\label{eq:mm_local_varrho}
\end{equation}
where $0<\epsilon\ll 1$ avoids division by zero. In~\eqref{eq:mm_local_varrho}  $m$ and $m'$ denote two distinct modalities among the $M$ utilized modalities (i.e.,  $\mathbf{g}_{n,m'}^{\mathrm{Sh}}(\mathbf{w})$, which is defined similar to~\eqref{eq:mm_shared_modality_gradient}, represents the modality-induced gradient of modality $m'$ on the shared fusion-decoder parameters of the model).
\end{definition}

The coefficient $\varrho_n(\mathbf{w})$ in~\eqref{eq:mm_local_varrho} measures whether the modalities induce
\textit{compatible} or \textit{conflicting} gradient descent directions on the shared fusion-decoder
parameters. Specifically, if $\varrho_n(\mathbf{w})>0$, the modality-induced shared gradients
are positively aligned and the modalities are locally complementary. If
$\varrho_n(\mathbf{w})<0$, the modality-induced shared gradients conflict and
the modalities are locally distractive. Note that 
when $M=1$, the pairwise summation in the numerator of
\eqref{eq:mm_local_varrho} is empty. Hence,
$
\varrho_n(\mathbf{w})=0$ when  $M=1$.
Thus, as expected, there is no cross-modal interaction in the single-modal setting. We next show the range of  $\varrho_n$.

\begin{lemma}[Boundedness of the Local Cross-Modal Interaction Coefficient]\label{lemma:11}
Let the local cross-modal interaction coefficient at AP $n$ be defined as~\eqref{eq:mm_local_varrho}. Then, for $M\geq 2$, we have
\begin{equation}
-\frac{1}{M-1}
<
\varrho_n(\mathbf{w})
<
1.
\label{eq:varrho_range_refined}
\end{equation}
Consequently,
\begin{equation}
-1
<
\varrho_n(\mathbf{w})
<
1.
\label{eq:varrho_range}
\end{equation}
\end{lemma}

\begin{proof}
See Appendix A.
\end{proof}

The coefficient $\varrho_n(\mathbf{w})$ in~\eqref{eq:mm_local_varrho}
characterizes cross-modal interaction within an individual AP. However, the
convergence behavior of FL is governed not only by local
modality interactions, but also by how these interactions behave after
aggregation across APs. In particular, even if modalities are complementary at
some APs, their induced shared gradients may become less aligned after
averaging across APs. Conversely, aggregation can also suppress
local modality conflicts and yield a more coherent global gradient descent direction.
To distinguish these two effects, we next define local-average and global
cross-modal interaction coefficients.

\begin{definition}[Global Cross-Modal Interaction and Average Local Cross-Modal Interaction]\label{Def:33}
Let $\overline{\mathbf{g}}_{m}^{\mathrm{Sh}}(\mathbf{w})$ denote the average modality-conditioned shared gradient associated
with modality $m$ across APs, defined as
\begin{equation}
\overline{\mathbf{g}}_{m}^{\mathrm{Sh}}(\mathbf{w})
\triangleq
\frac{1}{N}
\sum_{n=1}^{N}
\mathbf{g}_{n,m}^{\mathrm{Sh}}(\mathbf{w}).
\label{eq:mm_average_shared_gradient}
\end{equation}
We define the global cross-modal interaction coefficient as
\begin{equation}
\varrho_{\mathrm{Glob}}(\mathbf{w})
\triangleq
\frac{
2
\sum_{m=1}^{M-1} \sum_{m'=m+1}^{M}
\left\langle
\overline{\mathbf{g}}_{m}^{\mathrm{Sh}}(\mathbf{w}),
\overline{\mathbf{g}}_{m'}^{\mathrm{Sh}}(\mathbf{w})
\right\rangle
}{
(M-1)
\sum_{m=1}^{M}
\left\|
\overline{\mathbf{g}}_{m}^{\mathrm{Sh}}(\mathbf{w})
\right\|^2
+\epsilon
},
\label{eq:mm_global_varrho}
\end{equation}
where $0<\epsilon\ll 1$. Following the observations made in the proof of Lemma~\ref{lemma:11}, we have $-1 \leq \varrho_{\mathrm{Glob}}(\mathbf{w})\leq 1$. We also define the average local cross-modal
interaction coefficient as
\begin{equation}
\overline{\varrho}_{\mathrm{Loc}}(\mathbf{w})
\triangleq
\frac{1}{N}
\sum_{n=1}^{N}
\varrho_n(\mathbf{w}),
\label{eq:mm_average_local_varrho}
\end{equation}
where $\varrho_n(\mathbf{w})$ is given by~\eqref{eq:mm_local_varrho}. Moreover, revisiting the result of Lemma~\ref{lemma:11} implies that $-1 \leq \overline{\varrho}_{\mathrm{Loc}}(\mathbf{w})\leq 1$.
\end{definition}

In~\eqref{eq:mm_average_local_varrho}, the quantity $\overline{\varrho}_{\mathrm{Loc}}(\mathbf{w})$ measures the
average cross-modal compatibility within local AP objectives before
aggregation, while $\varrho_{\mathrm{Glob}}(\mathbf{w})$ in \eqref{eq:mm_global_varrho} measures the
cross-modal compatibility after averaging modality-conditioned shared gradients
across APs. Therefore, these two quantities distinguish whether
complementarity or distractiveness is primarily a local phenomenon or whether
it persists at the global federated level.
Specifically, when $\varrho_{\mathrm{Glob}}(\mathbf{w})>0$, the aggregated
modality-conditioned shared gradients are positively aligned, indicating that
the modalities remain complementary after aggregation. In contrast, when
$\varrho_{\mathrm{Glob}}(\mathbf{w})<0$, the aggregated modality-conditioned
shared gradients conflict with each other, indicating that the modalities
remain distractive at the global level.
Finally, we note that when $M=1$, the pairwise summations in
the numerators of \eqref{eq:mm_global_varrho} and~\eqref{eq:mm_local_varrho} vanish, and thus
$
\overline{\varrho}_{\mathrm{Loc}}(\mathbf{w})=
\varrho_{\mathrm{Glob}}(\mathbf{w})=0.
$
Therefore, as expected, in the single-modal case, there is no local or global cross-modal
interaction.



Next, to capture the joint impact of $\overline{\varrho}_{\mathrm{Loc}}(\mathbf{w})$ and
$\varrho_{\mathrm{Glob}}(\mathbf{w})$ in a compact form (i.e., to jointly capture whether modalities are locally complementary or distractive and
whether the aggregated multi-modal descent direction remains coherent at the
global level), we define the following
multi-modal interaction factor.

\begin{definition}[Multi-Modal Interaction Factor]
We define the  multi-modal interaction factor in our FL system as
\begin{equation}
\Gamma_{\mathrm{MM}}(\mathbf{w})
\triangleq
\frac{
1+\overline{\varrho}_{\mathrm{Loc}}(\mathbf{w})
}{
1+\varrho_{\mathrm{Glob}}(\mathbf{w})
}.
\label{eq:mm_interaction_factor}
\end{equation}
We further assume that this factor is uniformly bounded,
i.e.,
\begin{equation}
\Gamma_{\mathrm{MM}}(\mathbf{w})
\leq
\overline{\Gamma}_{\mathrm{MM}},
\quad \forall \mathbf{w}.
\label{eq:mm_interaction_factor_bound}
\end{equation}
\end{definition}

In~\eqref{eq:mm_interaction_factor}, the factor $\Gamma_{\mathrm{MM}}(\mathbf{w})$ captures how cross-modal
interactions affect the mismatch between local and global gradient descent directions. Specifically, the numerator $1+\overline{\varrho}_{\mathrm{Loc}}(\mathbf{w})$ reflects the
average strength of local cross-modal interaction, while the denominator
$1+\varrho_{\mathrm{Glob}}(\mathbf{w})$ reflects the strength of cross-modal
interaction after aggregation. Since
 $-1<\varrho_{\mathrm{Glob}}(\mathbf{w}),\overline{\varrho}_{\mathrm{Loc}}(\mathbf{w})<1$ as discussed in Definition~\ref{Def:33},
the shifted quantities
$1+\overline{\varrho}_{\mathrm{Loc}}(\mathbf{w})$ and
$1+\varrho_{\mathrm{Glob}}(\mathbf{w})$ are positive, making
$\Gamma_{\mathrm{MM}}(\mathbf{w})$ well-defined.
Moreover, when modalities remain complementary after aggregation,
$\varrho_{\mathrm{Glob}}(\mathbf{w})$ becomes positive, which increases the
denominator in~\eqref{eq:mm_interaction_factor} and reduces
$\Gamma_{\mathrm{MM}}(\mathbf{w})$. In this case, local multi-modal model updates
are more consistent with the global descent direction, resulting in an improved global model convergence, which will be further shown by our later convergence bounds. In contrast, when modalities
remain distractive after aggregation, $\varrho_{\mathrm{Glob}}(\mathbf{w})$
becomes negative, which decreases the denominator and increases
$\Gamma_{\mathrm{MM}}(\mathbf{w})$. This enlarges the mismatch between local
and global gradients and slows the global model convergence.
Finally, we note that when $M=1$ (i.e., in single-modal scenario), as expected,  we have
$
\overline{\varrho}_{\mathrm{Loc}}(\mathbf{w})=0,
\qquad
\varrho_{\mathrm{Glob}}(\mathbf{w})=0.
$
Therefore,
$
\Gamma_{\mathrm{MM}}(\mathbf{w})
=
\frac{1+0}{1+0}
=
1.
$

The above discussion shows that multi-modal interactions alter the mismatch
between local and global gradients through the factor ${\Gamma}_{\mathrm{MM}}$, and thus its bound
$\overline{\Gamma}_{\mathrm{MM}}$. We now incorporate this effect into the
\textit{gradient diversity} model, which captures the \textit{data heterogeneity} across the APs. Specifically, in the single-modal setting, gradient diversity across the APs is usually
controlled by bounding the average squared norm of local gradients in terms of
the squared norm of the global gradient and an additive heterogeneity term (e.g., see \textit{Assumption 3} in \cite{wang2020tackling}).
In the multi-modal setting, this drift is affected not only by AP-level data
heterogeneity, but also by whether modality-induced shared gradients are
complementary or distractive across APs. This motivates our following
assumption.

\begin{assumption}[Gradient Diversity APs]\label{Assump:1}
There exist finite constants $\zeta_1^{\mathrm{AP}}\geq 1$ and
$\zeta_2^{\mathrm{AP}}\geq 0$ such that
\begin{equation}
\hspace{-2mm}
\resizebox{0.92\linewidth}{!}{$
\frac{1}{N}
\sum_{n=1}^{N}
\left\|
\nabla f_n(\mathbf{w})
\right\|^2
\leq
\zeta_1^{\mathrm{AP}}
\overline{\Gamma}_{\mathrm{MM}}
\left\|
\nabla f(\mathbf{w})
\right\|^2
+
\zeta_2^{\mathrm{AP}},
~ \forall \mathbf{w}.$}\hspace{-2mm}
\label{eq:mm_ap_drift}
\end{equation}
\end{assumption}

The constants $\zeta_1^{\mathrm{AP}}$ and $\zeta_2^{\mathrm{AP}}$ capture the
standard AP-level data heterogeneity, recovering \textit{Assumption 3} in \cite{wang2020tackling} on data heterogeneity in single-modal FL. Specifically, $\zeta_1^{\mathrm{AP}}$
controls the multiplicative mismatch between local and global gradients, while
$\zeta_2^{\mathrm{AP}}$ captures residual gradient disagreement that cannot be
explained purely by the global gradient norm. The term
$\overline{\Gamma}_{\mathrm{MM}}$ introduces the additional effect of
multi-modal interactions. Therefore, the effective  coefficient in the
multi-modal setting is
$
\zeta_1^{\mathrm{AP}}
\overline{\Gamma}_{\mathrm{MM}}.
$ 
This effective coefficient directly determines how strongly the local-gradient
dispersion affects the convergence bound. When the modalities are
complementary and their shared-gradient directions remain aligned after
aggregation, $\overline{\Gamma}_{\mathrm{MM}}$ becomes smaller, which reduces
$\zeta_1^{\mathrm{AP}}
\overline{\Gamma}_{\mathrm{MM}}$  (i.e., it implies a smaller gradient diversity across APs and thus a less heterogeneous data distribution). In contrast,
when the modalities are distractive and their shared-gradient directions
conflict locally or globally, $\overline{\Gamma}_{\mathrm{MM}}$ becomes
larger, which increases $\zeta_{1,\mathrm{eff}}$ (i.e., it implies a larger gradient diversity across APs and thus a more heterogeneous data distribution). We will revisit these nuances when explaining our convergence bound in our subsequent discussions. Finally, we note that when $M=1$, as discussed above, we have $\Gamma_{\mathrm{MM}}(\mathbf{w})
=\overline{\Gamma}_{\mathrm{MM}}=
1,$  and thus \eqref{eq:mm_ap_drift} reduces to
$
\frac{1}{N}
\sum_{n=1}^{N}
\left\|
\nabla f_n(\mathbf{w})
\right\|^2
\leq
\zeta_1^{\mathrm{AP}}
\left\|
\nabla f(\mathbf{w})
\right\|^2
+
\zeta_2^{\mathrm{AP}},
\quad \forall \mathbf{w}.
$
This is the standard AP-drift condition used in the single-modal convergence
analysis (i.e., \textit{Assumption 3} in \cite{wang2020tackling}). 

We next revisit the stochastic-gradient noise assumption under the introduced
multi-modal architecture. In doing so, we note that in the  single-modal analysis, stochasticity
is captured by a scalar variance bound on the mini-batch gradient noise. In the
multi-modal setting, however, the stochastic gradient contains noise from two
different sources: (i) modality-specific encoder blocks and (ii) the shared
fusion-decoder block. The second source is especially important because
different modalities jointly affect the same shared parameters. Therefore, the
noise terms induced by different modalities on the shared fusion-decoder
parameters may be correlated, and these correlations can either increase or
decrease the effective stochastic-gradient variance. This motivates our
following assumption.

\begin{assumption}[Multi-Modal Stochastic-Gradient Noise]\label{Assump:2}
At each AP $n$,  the stochastic
gradient computed from a mini-batch of multi-modal samples, i.e., $\widetilde{\nabla} f_n(\mathbf{w})$, 
is an unbiased estimator of the full local multi-modal gradient, which implies
$
\mathbb{E}
\left[
\widetilde{\nabla} f_n(\mathbf{w})
\right]
=
\nabla f_n(\mathbf{w}),
\quad \forall n,\mathbf{w}.
$
Equivalently, defining the multi-modal stochastic-gradient noise as
$
\boldsymbol{\epsilon}_{n,\mathrm{MM}}(\mathbf{w})
\triangleq
\widetilde{\nabla} f_n(\mathbf{w})
-
\nabla f_n(\mathbf{w}),
$
we have
$
\mathbb{E}
\left[
\boldsymbol{\epsilon}_{n,\mathrm{MM}}(\mathbf{w})
\right]
=
\mathbf{0}.
$ Also, since the parameter vector consists of modality-specific encoder parameters
and shared fusion-decoder parameters, the noise vector admits the block
decomposition
\begin{equation}
\boldsymbol{\epsilon}_{n,\mathrm{MM}}(\mathbf{w})
=
\left[
\boldsymbol{\epsilon}_{n,1}^{\mathrm{Enc}}(\mathbf{w}),
\ldots,
\boldsymbol{\epsilon}_{n,M}^{\mathrm{Enc}}(\mathbf{w}),
\boldsymbol{\epsilon}_{n}^{\mathrm{Sh}}(\mathbf{w})
\right],
\label{eq:mm_noise_block_decomposition}
\end{equation}
where $\boldsymbol{\epsilon}_{n,m}^{\mathrm{Enc}}(\mathbf{w})$ denotes the
stochastic-gradient noise on the encoder of modality $m$, and
$\boldsymbol{\epsilon}_{n}^{\mathrm{Sh}}(\mathbf{w})$ denotes the
stochastic-gradient noise on the shared fusion-decoder parameters. Further, the shared noise can be decomposed into modality-induced components as
$
\boldsymbol{\epsilon}_{n}^{\mathrm{Sh}}(\mathbf{w})
=
\sum_{m=1}^{M}
\boldsymbol{\epsilon}_{n,m}^{\mathrm{Sh}}(\mathbf{w}),
$
where $\boldsymbol{\epsilon}_{n,m}^{\mathrm{Sh}}(\mathbf{w})$ denotes the
noise contribution induced by modality $m$ on the shared fusion-decoder
parameters.
Using this decomposition, the effective multi-modal stochastic-gradient
variance at AP $n$ is 
\begin{equation}
\hspace{-2mm}
\resizebox{0.92\linewidth}{!}{$
\begin{aligned}
\sigma_{n,\mathrm{MM}}^2(\mathbf{w})
\triangleq
&
\sum_{m=1}^{M}
\mathbb{E}
\left[
\left\|
\boldsymbol{\epsilon}_{n,m}^{\mathrm{Enc}}(\mathbf{w})
\right\|^2
\right]
+
\sum_{m=1}^{M}
\mathbb{E}
\left[
\left\|
\boldsymbol{\epsilon}_{n,m}^{\mathrm{Sh}}(\mathbf{w})
\right\|^2
\right]
\\
&\quad
+
2 \sum_{m=1}^{M-1} \sum_{m'=m+1}^{M}
\mathbb{E}
\left[
\left\langle
\boldsymbol{\epsilon}_{n,m}^{\mathrm{Sh}}(\mathbf{w}),
\boldsymbol{\epsilon}_{n,m'}^{\mathrm{Sh}}(\mathbf{w})
\right\rangle
\right].
\end{aligned}
$}
\hspace{-2mm}
\label{eq:mm_effective_noise_variance}
\end{equation}
We assume that this effective variance is uniformly bounded, i.e.,
\begin{equation}
\mathbb{E}
\left[
\left\|
\boldsymbol{\epsilon}_{n,\mathrm{MM}}(\mathbf{w})
\right\|^2
\right]
=
\sigma_{n,\mathrm{MM}}^2(\mathbf{w})
\leq
\sigma_{\mathrm{MM}}^2,
\quad \forall n,\mathbf{w},
\label{eq:mm_noise_uniform_bound}
\end{equation}
where
\begin{equation}
\sigma_{\mathrm{MM}}^2
\triangleq
\sup_{n\in\mathcal{N},\,\mathbf{w}}
\sigma_{n,\mathrm{MM}}^2(\mathbf{w}).
\label{eq:mm_global_noise_bound}
\end{equation}
\end{assumption}

The definition in~\eqref{eq:mm_effective_noise_variance}, which is then encoded in the upperbound  in~\eqref{eq:mm_global_noise_bound}, explicitly
captures the effect of multi-modality through three terms. The first term (i.e., $\sum_{m=1}^{M}
\mathbb{E}
\left[
\left\|
\boldsymbol{\epsilon}_{n,m}^{\mathrm{Enc}}(\mathbf{w})
\right\|^2
\right]$)
captures stochasticity in the modality-specific encoder blocks. The second
term (i.e., $\sum_{m=1}^{M}
\mathbb{E}
\left[
\left\|
\boldsymbol{\epsilon}_{n,m}^{\mathrm{Sh}}(\mathbf{w})
\right\|^2
\right]$) captures the individual modality-induced stochasticity on the shared
fusion-decoder block. The third term (i.e., $2
\sum_{m=1}^{M-1} \sum_{m'=m+1}^{M}
\mathbb{E}
\left[
\left\langle
\boldsymbol{\epsilon}_{n,m}^{\mathrm{Sh}}(\mathbf{w}),
\boldsymbol{\epsilon}_{n,m'}^{\mathrm{Sh}}(\mathbf{w})
\right\rangle
\right]$) captures cross-modal stochastic
interactions on the shared fusion-decoder block through the inner products
\begin{equation}
\mathbb{E}
\left[
\left\langle
\boldsymbol{\epsilon}_{n,m}^{\mathrm{Sh}}(\mathbf{w}),
\boldsymbol{\epsilon}_{n,m'}^{\mathrm{Sh}}(\mathbf{w})
\right\rangle
\right].
\label{eq:mm_noise_inner_product}
\end{equation}
If these inner products are positive, then the modality-induced noise
components reinforce each other on the shared parameters, increasing
$\sigma_{\mathrm{MM}}^2$ and worsening the convergence performance. If they are
small or negative, then the modality-induced noise components are weakly
correlated or partially cancel each other, reducing the effective noise variance.

Finally, we note that when $M=1$, there is only one encoder block and one modality-induced shared
noise component. Hence, the pairwise cross-modal noise summation in
\eqref{eq:mm_effective_noise_variance} vanishes, and we obtain
\begin{equation}
\sigma_{n,\mathrm{MM}}^2(\mathbf{w})
=
\mathbb{E}
\left[
\left\|
\boldsymbol{\epsilon}_{n,1}^{\mathrm{Enc}}(\mathbf{w})
\right\|^2
\right]
+
\mathbb{E}
\left[
\left\|
\boldsymbol{\epsilon}_{n,1}^{\mathrm{Sh}}(\mathbf{w})
\right\|^2
\right].
\label{eq:single_modal_noise}
\end{equation}
The right-hand side is exactly the stochastic-gradient variance of the
single-modal model, which can be denoted by $\sigma^2$. Therefore,
$
\sigma_{\mathrm{MM}}^2
=
\sigma^2
$
and the Assumption in~\eqref{eq:mm_noise_uniform_bound} reduces to the standard
bounded SGD noise assumption in the single-modal setting (e.g., \textit{Assumption 2} in \cite{wang2020tackling}).

We next present our final assumption for convergence analysis, which is a standard assumption in the literature.
\begin{assumption}[Smoothness of Local Loss Functions]\label{Assump:3}
    The local objective function $f_{n}(\mathbf{w})$ is $L$-Lipschitz smooth such that $f_{n}(\mathbf{w}') \leq f_{n}(\mathbf{w}) + \langle \nabla f_{n}(\mathbf{w}), \mathbf{w}'-\mathbf{w}\rangle + L/2 ||\mathbf{w}'-\mathbf{w}||^{2}$ is satisfied for any two parameters $\mathbf{w}$ and $\mathbf{w}'$.
\end{assumption}

We next present our main result, which represents the global model convergence in our multi-modal setting of interest.

\begin{theorem}[Performance of the Trained Global Model] \label{thm2}
Considering the above assumptions and definitions, and $f^{\star} = \text{min}_{\mathbf{w}} f(\mathbf{w})$ while choosing the step size to satisfy $\eta \leq \min\{\frac{1}{JL}, \frac{1}{2L} \sqrt{\frac{\Omega }{(\zeta_1^{\mathrm{AP}} \overline{\Gamma}_{\mathrm{MM}}+\Omega )J(J-1)}}\}$ with constant $\Omega<1$, the convergence performance of Algorithm 1 is given by
\begin{equation}\label{mainTh22}
\begin{aligned}
    &\frac{1}{T} \sum_{t=0}^{T-1} \mathbb{E} \Big[ \left\Vert \nabla f(\mathbf{w}^{(t)})\right\Vert^2 \Big]\leq 
     \frac{f(\mathbf{w}^{(0)})-f^{\star}}{T \eta J \frac{1}{2} (1-\Omega)}\\
&+\frac{ L^2
        {2 \eta^2 (J-1) \sigma_{\mathrm{MM}}^2 }}{  (1-\Omega)} +\frac{L}{(1-\Omega)}\eta\sigma_{\mathrm{MM}}^2
        \\&
        + \frac{ L^2 4\eta^2 J(J-1)2 \zeta_2^{\mathrm{AP}}}{  (1-\Omega)}
   \end{aligned}
\end{equation} 
Accordingly, choosing $\eta=\frac{\varphi}{\sqrt{TJ}}$, where $\varphi$ is chosen small enough such that $\frac{\varphi}{\sqrt{TJ}}\leq \min\{\frac{1}{JL}, \frac{1}{2L} \sqrt{\frac{\Omega }{(\zeta_1^{\mathrm{AP}} \overline{\Gamma}_{\mathrm{MM}}+\Omega )J(J-1)}}\}$, the bound in~\eqref{mainTh22} implies that
$
\frac{1}{T}\sum_{t=0}^{T-1} \|\nabla f(\mathbf{w}^{(t)})\|^{2}
\leq \mathcal{O}\!\left(\frac{1}{\sqrt{T}}\right).
$
Therefore, we have
$
\lim_{T\rightarrow\infty}\frac{1}{T}\sum_{t=0}^{T-1}\|\nabla f(\mathbf{w}^{(t)})\|^{2}=0,
$
which indicates that the algorithm converges to a stationary point as $T\rightarrow\infty$.
\end{theorem}
\begin{proof} See Appendix B. \end{proof}

\noindent \textbf{Discussion of the Results.}
Theorem~\ref{thm2} shows that the proposed algorithm preserves the standard SGD convergence rate of
$\mathcal{O}(1/\sqrt{T})$  to a stationary point, while the
constants in the bound and the condition on the step size explicitly reveal the effect of multi-modality. In
particular, 
multi-modality affects the convergence behavior through two coupled channels:
(i) the alignment between local and global descent directions, captured by
$\overline{\Gamma}_{\mathrm{MM}}$, and (ii) the stochastic variability of the
multi-modal mini-batch gradient, captured by $\sigma_{\mathrm{MM}}^2$.
Specifically, the step size condition in Theorem~\ref{thm2}  clarifies the role of
multi-modal complementarity and distractiveness. This is because the step size
must satisfy
$
\eta
\leq
\frac{1}{2L}
\sqrt{
\frac{\Omega}{
\left(
\zeta_1^{\mathrm{AP}}\overline{\Gamma}_{\mathrm{MM}}
+
\Omega
\right)
J(J-1)
}
}.
$
Thus, when the modalities are complementary, the modality-induced shared
gradients are better aligned locally and globally, which leads to a smaller
$\overline{\Gamma}_{\mathrm{MM}}$. This reduces 
$\zeta_1^{\mathrm{AP}}\overline{\Gamma}_{\mathrm{MM}}$ and allows a less
restrictive step-size condition. In contrast, when the modalities are
distractive, their induced shared gradients conflict on the fusion-decoder
parameters, which increases $\overline{\Gamma}_{\mathrm{MM}}$. This makes the
step-size condition more restrictive and amplifies the local-update drift
caused by performing $J$ local SGD steps before model aggregation. Next, the noise-dependent terms in~\eqref{mainTh22} (i.e., $
\frac{
L^2 2\eta^2 (J-1)\sigma_{\mathrm{MM}}^2
}{
1-\Omega
}
+
\frac{
L\eta\sigma_{\mathrm{MM}}^2
}{
1-\Omega
}
$) show the second way in which
multi-modality affects convergence. Specifically, the effective variance $\sigma_{\mathrm{MM}}^2$, given by \eqref{eq:mm_global_noise_bound}, contains the stochastic noise
from modality-specific encoders, the modality-induced noise on the shared
fusion-decoder parameters, and the cross-modal noise correlations on the
shared parameters. If the stochastic perturbations induced by different
modalities on the shared parameters are weakly correlated or partially
canceling, then $\sigma_{\mathrm{MM}}^2$ becomes smaller, improving the
convergence constant. Conversely, if these perturbations are positively
correlated and reinforce each other, then $\sigma_{\mathrm{MM}}^2$ becomes
larger, increasing the noise floor in the bound.

Further, the dependence on $J$ in~\eqref{mainTh22} also has an intuitive interpretation. Increasing the
number of local SGD steps can reduce communication frequency, but it also
increases the discrepancy between the locally updated AP models and the global
model due to data heterogeneity across the APs. This effect appears through the factor $J(J-1)$ in the step-size
condition and in the drift-related term of the convergence bound (i.e., $\frac{ L^2 4\eta^2 J(J-1)2 \zeta_2^{\mathrm{AP}}}{  (1-\Omega)}$). Therefore,
when AP-level data heterogeneity or modality distractiveness is strong, a smaller
step size or fewer local steps may be needed to prevent local model updates from
drifting away from each other (i.e., to avoid local model bias).

It is important to note that, as expected, Theorem~\ref{thm2} indicates that multi-modality does not automatically
benefit convergence. Specifically, complementary modalities improve the
bound by reducing the coefficient
$\zeta_1^{\mathrm{AP}}\overline{\Gamma}_{\mathrm{MM}}$ and potentially lowering
$\sigma_{\mathrm{MM}}^2$ through weakly correlated stochastic perturbations. However, distractive modalities have the opposite effect: they enlarge
$\overline{\Gamma}_{\mathrm{MM}}$, impose a more conservative step-size
condition, and may increase $\sigma_{\mathrm{MM}}^2$ if their stochastic
effects on the shared fusion-decoder parameters are aligned. Hence, the
theorem makes explicit that the convergence behavior of federated multi-modal
learning depends not merely on the number of modalities, but on whether their
shared-gradient directions and stochastic perturbations are complementary or
conflicting. This interpretation also motivates the choice of Modality 1, Modality 2, and Modality 3 (introduced in Sec. \ref{mm_rep}). In particular, in our considered FLAME architecture, the three introduced
modalities are expected to provide largely aligned and complementary
information for the target classification task. Therefore, their
modality-induced shared gradients should exhibit positive cross-modal
interaction, resulting in a smaller 
$\overline{\Gamma}_{\mathrm{MM}}$ and a more favorable model convergence.
Accordingly, in the subsequent section, we empirically show that jointly using
all three modalities improves the convergence behavior compared with using
individual modalities or less informative modality combinations. This confirms
the above theoretical insight that multi-modal FL benefits most when the available
modalities induce compatible gradient descent directions and do not introduce strongly
conflicting stochastic perturbations on the shared fusion-decoder parameters.

\begin{figure*}[ht]
    \centering
    \includegraphics[width=\textwidth]{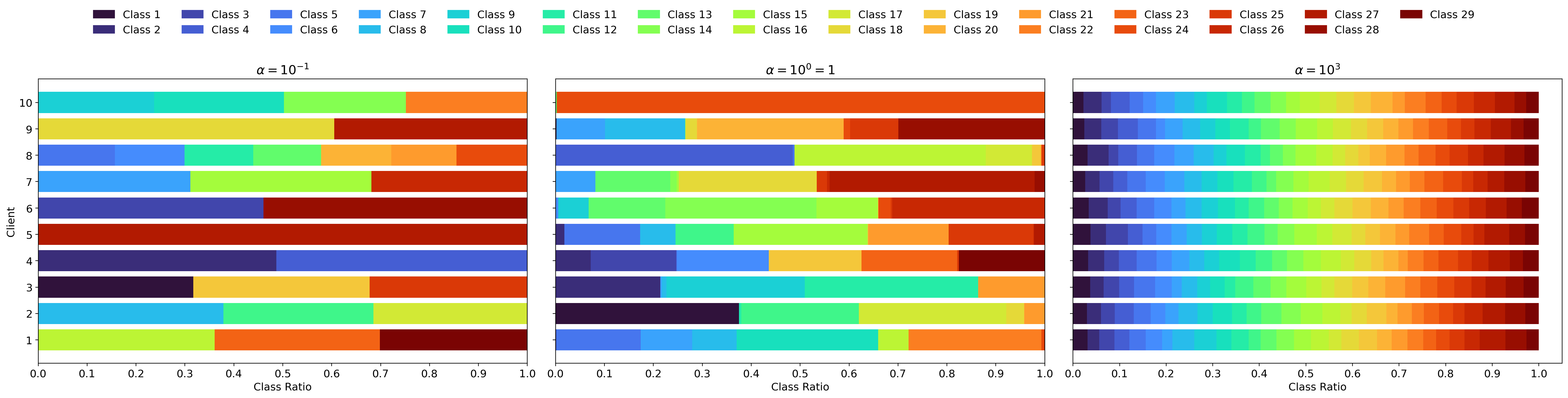}
    \caption{Data distribution across classes for $10$ clients in extreme (left), moderate (middle), and low (right) data heterogeneity scenarios.}
    \label{fig:heterogeneity}
\end{figure*}

\section{Performance Evaluation} \label{results}

In this section, we first describe our empirical setup including our employed datasets (Sec.~\ref{datasets}) as well as our FL architecture (Sec.~\ref{exp_setup}). Afterwards, we (i) compare our method against other modality combinations in Sec.~\ref{sec:cross-combination}; (ii) perform experiments to verify the consistency of the performance gain of our method across the existing FL baselines in Sec.~\ref{sec:baselines}; (iii) perform ablation studies on the number of clients (Sec.~\ref{sec:ablation-client}), number of local training rounds (Sec.~\ref{sec:ablation-local-rounds}), and datasets used to perform the experiments (Sec.~\ref{sec:oracle-ablation}); (iv) perform a set of experiments to investigate the perception of our trained model towards the multi-modal data in Sec.~\ref{sec:tsne}; (v) conduct a set of experiments to isolate the impact of model scale on performance in Sec.~\ref{sec:capacity-match}; (vi) perform a set of experiments to investigate the computation and communication overhead of model training in Sec.~\ref{sec:comm-comp-overheads}.
All implementations are available at our public repository at \href{https://github.com/KasraBorazjani/flame}{https://github.com/KasraBorazjani/flame}.

\subsection{Datasets} \label{datasets}



We perform our experiments on two datasets: 1) the ORBIT Testbed WiFi Transmitter Dataset~\cite{dataset}, which we perform our original analysis and ablation studies on it and 2) the ORACLE dataset~\cite{sankhe2019oracle}, which we use to repeat our main analyses to check for the generalizability of our results.
The ORBIT  dataset  consists of $163$ WiFi modules as transmitters and a software defined radio (USRP N210) as a receiver. The transmitted data has a center frequency of 2462 MHz and a bandwidth of 20 MHz. In addition, the received waveforms used for training consist of $\ell = 256$ I/Q samples, capturing the received signal's preamble, which can be used in place of the entire received payload to perform effective device RF fingerprinting without incurring unnecessary training overhead \cite{preamble1,preamble2}.
The ORACLE dataset is composed of $16$ USRP X310 software-defined radios, serving as transmitters, and a fixed USRP B210 as the receiver. Various distance settings between the transmitters and the receiver have been recorded in the dataset, from which we use the data gathered at the maximum distance of $62$ft to train our model. The transmitted signals possess a center frequency of 2450 MHz with a sampling rate of 5 MS/s. The received waveforms used for training consist of $\ell = 256$ I/Q samples per segment for compatibility with the experiments performed on the first dataset.
The preprocessing for each of the datasets is explained in full detail in Appendix C.

\subsection{Experimental Setup} \label{exp_setup}
We distribute the data based on the Dirichlet distribution $\mathsf{Dir}(\alpha)$ \cite{hsu2019measuring} of the class labels $\mathcal{A}_i$ (corresponding to the transmitter that each datapoint is received from). The parameter $\alpha$ resembles the data homogeneity/heterogeneity of the synthesized local datasets; i.e., higher $\alpha$ will lead to higher homogeneity (lower data heterogeneity) and lower $\alpha$ will lead to lower homogeneity (higher data heterogeneity). Samples of data distribution across clients are visualized in Fig.~\ref{fig:heterogeneity} for different values of $\alpha$.
Given that the lowest level of data heterogeneity (depicted in Fig.~\ref{fig:heterogeneity}-right) is less likely to occur in a real-world scenario, we include $2$ values for $\alpha$ to account for different levels of data heterogeneity in the majority of our subsequent experiments: (i) $\alpha=1$ for moderate data heterogeneity, and (ii) $\alpha=0.1$ for extreme data heterogeneity.

We have taken a late-fusion multi-modal approach into consideration which is realized through an encoder-decoder architecture for our ML model, where a modality-specific encoder is considered for each input modality. After passing through the encoders, all available modalities' features are passed through a fully connected layer and then passed to the classification head for class prediction.

\begin{figure}[ht]
    \centering
    \includegraphics[width=0.49\textwidth, trim = 0 0 0 24, clip]{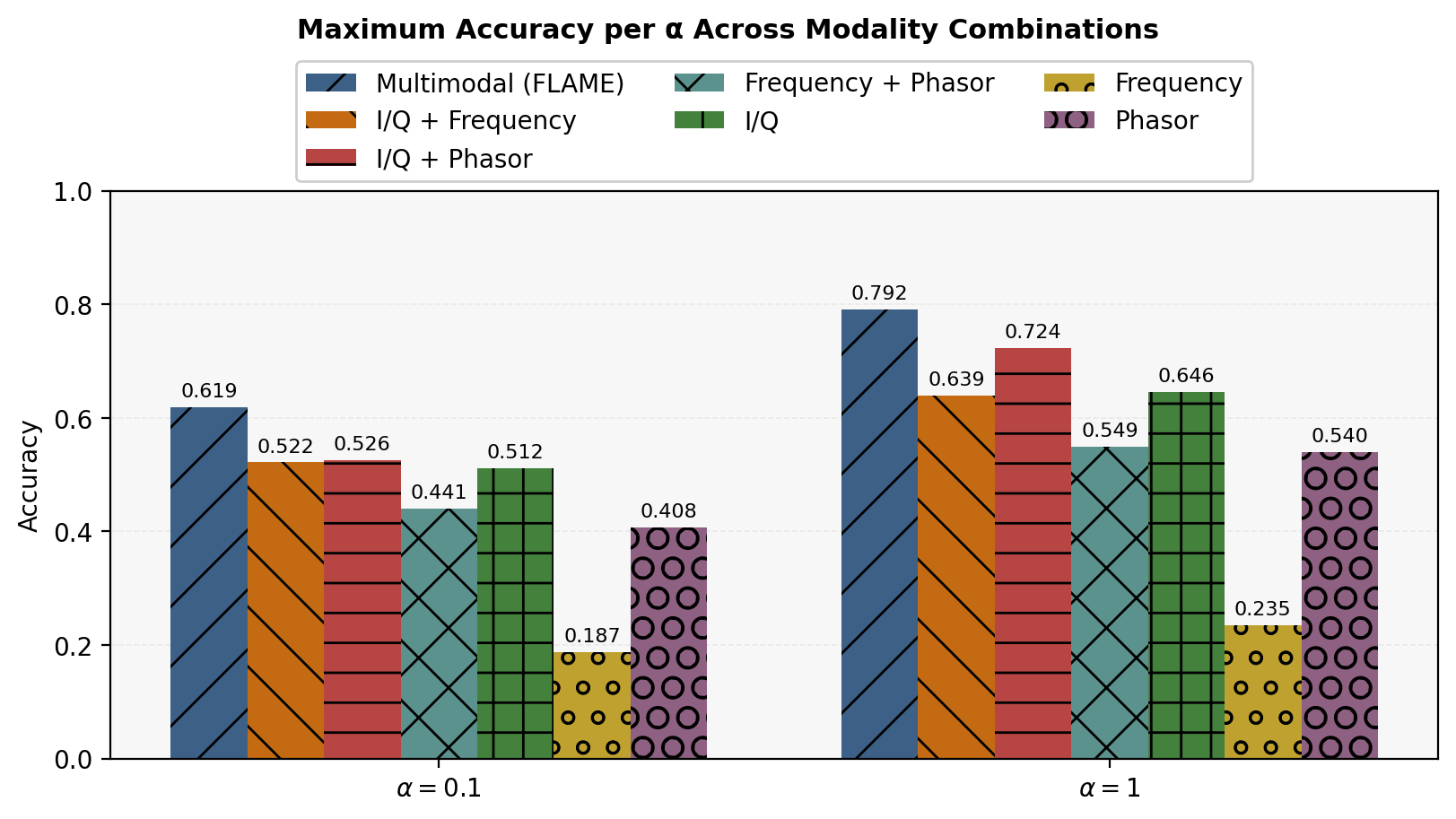}
    \caption{Maximum accuracies reached for $N=10$ clients under extreme ($\alpha=0.1$) and moderate ($\alpha=1$) data heterogeneities for all modality combinations.}
    \label{fig:cross-combination-bp}
\end{figure}

\begin{figure}[ht]
    \centering
    \includegraphics[width=0.97\linewidth, trim = 0 0 0 18, clip]{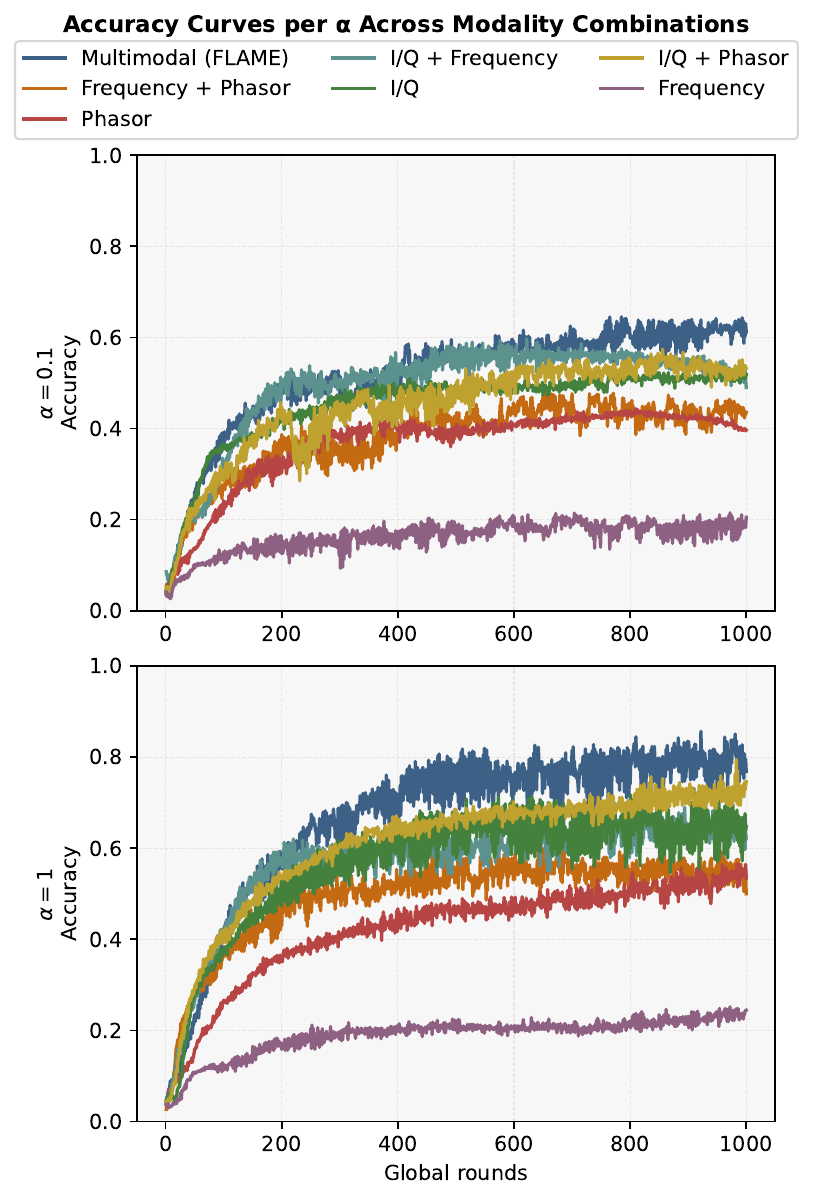}
    \caption{Accuracy curves for $N=10$ clients under extreme ($\alpha=0.1$) and moderate ($\alpha=1$) data heterogeneities for all modality combinations.}
    \label{fig:cross-combination-ac}
\end{figure}

We performed our simulations using Python Tensorflow library on NVIDIA A100 GPUs. We  experiment with various numbers of clients $N\in\{10,20,40\}$, numbers of local SGD iterations per aggregation round $J\in\{10, 20\}$, local step size of $\eta=0.0001$, batch size $B=128$, and number of global aggregation rounds $T=1000$. 
For comparison across different FL methods (as will be explained in Sec.~\ref{sec:baselines}), we have chosen FedAvg~\cite{FedAvg}, FedProx~\cite{fL_heterogeneous}, and SCAFFOLD~\cite{scaffold} as our baselines. We compare the performance of our proposed multi-modal approach against using only the I/Q modality with both encoder-based ML models and residual networks (using early fusion as pointed in \cite{mm_fingerprinting_dl}, unlike our method that considers late fusion) as baselines.
We use the Adam optimizer for training local models in FedAvg and FedProx. However, since the Adam optimizer's calculations coincide with the client weights calculated in SCAFFOLD, we use the SGD optimizer in its corresponding experiments with the tuned learning rate of $\eta=0.01$.

\subsection{Comparison Across Modality Combinations} \label{sec:cross-combination}
To study the capabilities of the multi-modal approach, we perform experiments on different modality combinations available in our dataset. We conduct these experiments for $N=10$ clients, each performing $J=10$ local SGD steps per aggregation round, and under two levels of data heterogeneity $\alpha\in\{0.1,1\}$. The results of our experiments are shown in Figs.~\ref{fig:cross-combination-bp} and \ref{fig:cross-combination-ac}.

As shown in Fig.~\ref{fig:cross-combination-bp}, which reflects the maximum value of the accuracy trends in Fig. 5, our method shows significant improvement compared to other modality combinations across different data heterogeneity scenarios: our multi-modal approach shows $9.3\%$ (for $\alpha=0.1$) and $6.8\%$ (for $\alpha=1$) improvement compared to the second-best modality combination in each data heterogeneity scenario.

\subsection{Comparison with Existing Baselines}\label{sec:baselines}

\begin{figure}[ht]
    \centering
    \includegraphics[width=\linewidth, trim = 0 0 0 0, clip]{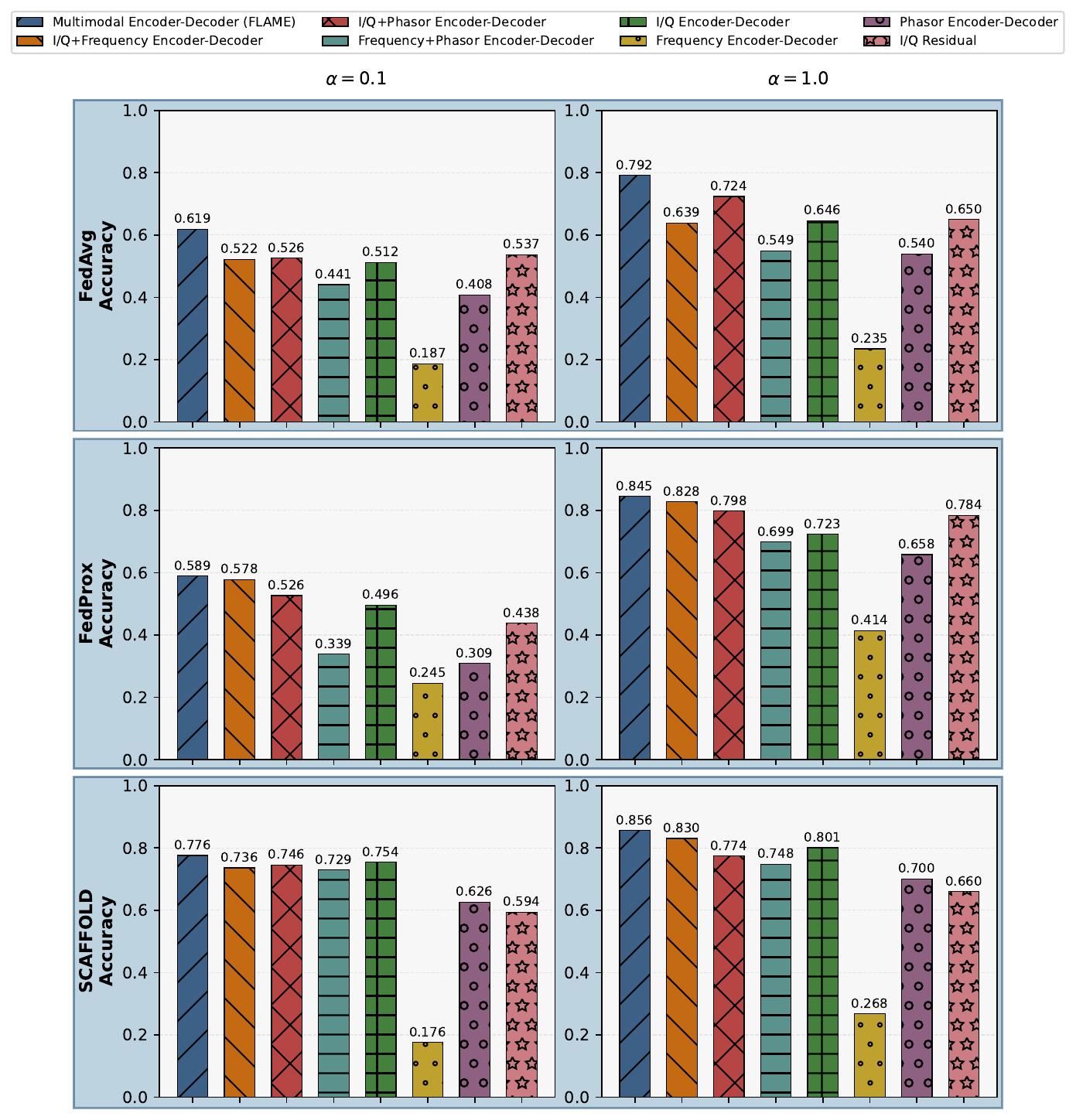}
    \caption{Performance plots across baselines and (left) extreme and (right) moderate data heterogeneity levels when FLAME is integrated into existing FL approaches. Methods are separated by boxes and modality combinations alongside the ML model architecture used for each bar are mentioned in the legend.}
    \label{fig:cross-method}
\end{figure}

We next perform a set of experiments to verify the consistency of the performance gain obtained by our proposed multi-modal approach across existing baselines in FL for which we consider FedAvg~\cite{FedAvg}, FedProx~\cite{fL_heterogeneous}, and SCAFFOLD~\cite{scaffold}. For each baseline, we compare the maximum accuracy reached by our proposed multi-modal approach (i.e., FLAME) using encoder-based ML models against that of other modality combinations and also the current trend in federated RF fingerprinting using I/Q modality with a residual ML model to also account for the effect of model architecture.  The results of our experiments are shown in Fig.~\ref{fig:cross-method}. We have tested the performance of each baseline method under various values for the learning rate ($\eta \in \{0.005, 0.001, 0.0005, 0.0001\}$ for FedAvg and FedProx and $\eta \in \{0.005, 0.001, 0.0005, 0.0001\}$ for SCAFFOLD) and reported the best performance for each method. It is important to note that the objective of this comparison is not to assess the relative superiority of different FL optimization methods, but rather to evaluate \textit{the impact of our proposed multi-modal processing under each method}. Therefore, comparisons are made within each FL algorithm, where the multi-modal configuration is contrasted against its counterparts of other modality and ML model configurations.



As shown in Fig.~\ref{fig:cross-method}, our proposed multi-modal approach outperforms all other approaches regardless of their ML model architecture across all baseline FL methods. The consistent observation of the performance gain obtained by our proposed multi-modal approach across all FL methods and both data heterogeneity levels demonstrates that FLAME can be used to conduct high-performance distributed RF fingerprinting regardless of the underlying used FL method. This consistency further indicates that the observed improvements stem from the complementary signal representations introduced by our multi-modal design rather than from algorithm-specific optimization effects.

\subsection{Ablation on the Number of Clients} \label{sec:ablation-client}
\begin{figure}[ht]
    \centering
    \includegraphics[width=\linewidth, trim = 0 0 0 19, clip]{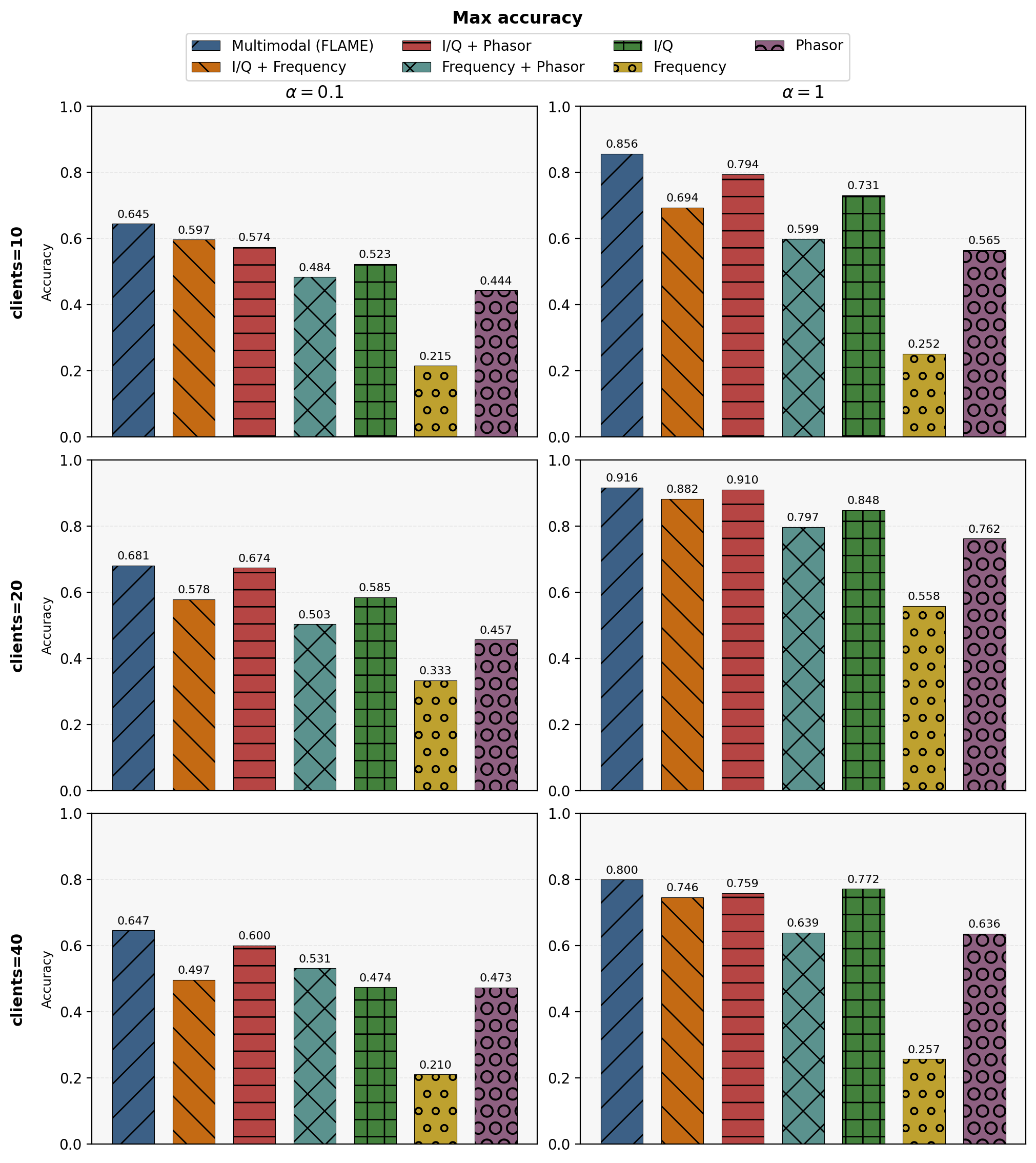}
    \caption{Maximum accuracy for different number of clients in the system ($N$) across combinations and data heterogeneity levels. Left column corresponds to extreme data heterogeneity and right column corresponds to moderate data heterogeneity. From top to bottom, the number of clients increases ($10$, $20$, and $40$, respectively).}
    \label{fig:client-ablation}
\end{figure}

To further examine the capabilities of our proposed approach, and the effect of different system sizes in each level of data heterogeneity, we performed a set of experiments under varying numbers of clients $N\in\{10, 20, 40\}$. We conducted $J=10$ SGD iterations per global aggregation round for all data heterogeneity scenarios. The best results for all modality combinations across various learning rates $\eta\in\{0.005, 0.001, 0.0005, 0.0001\}$ are included in Fig.~\ref{fig:client-ablation}. As shown in the figure, our multi-modal approach reaches the best performance across modality combinations and numbers of clients. Further, it maintains its superior performance across the considered data heterogeneity levels $\alpha\in\{0.1, 1\}$.

\subsection{Ablation on the Number of Local Rounds} \label{sec:ablation-local-rounds}
\begin{figure}[ht]
    \centering
    \includegraphics[width=\linewidth, trim = 0 0 0 19, clip]{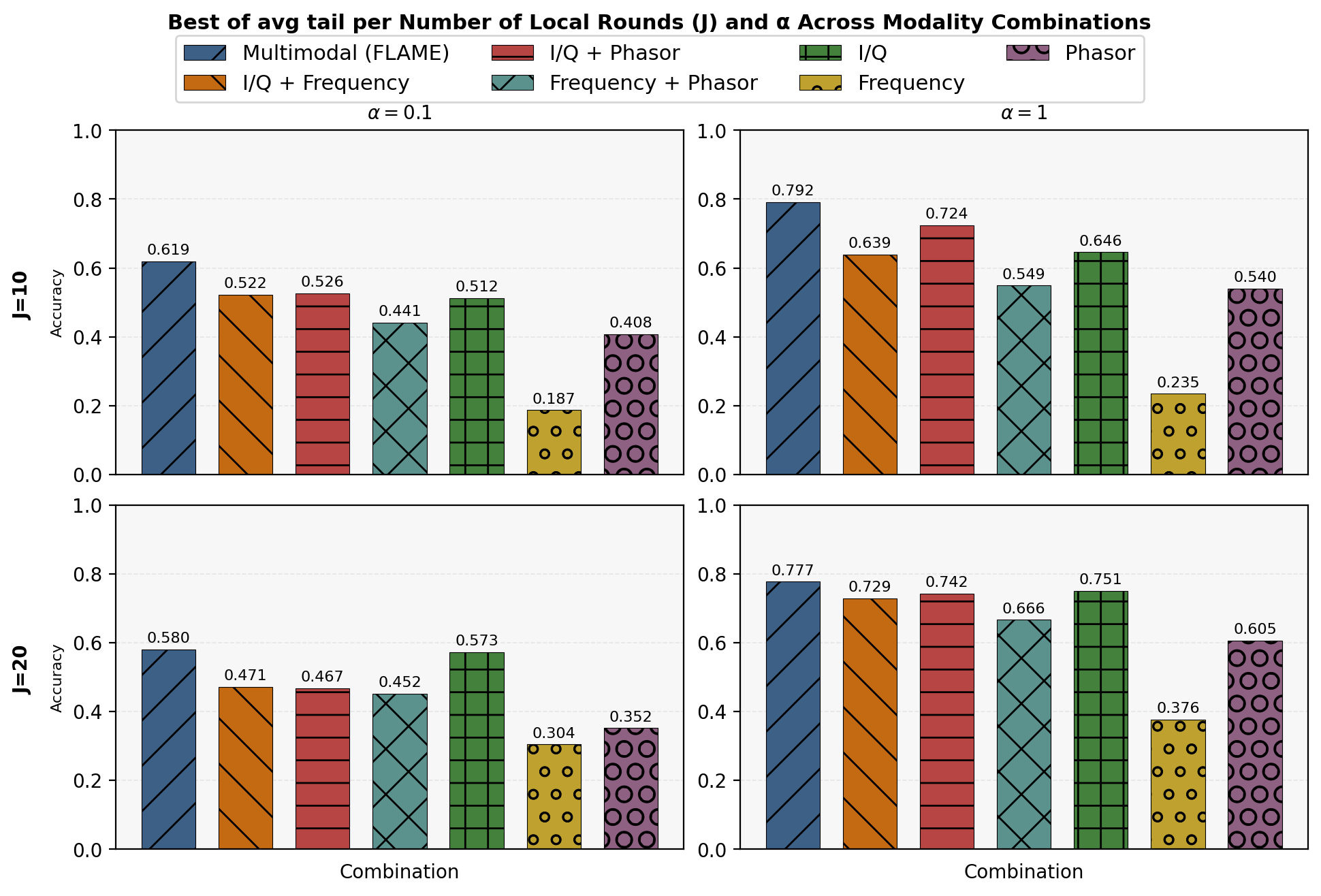}
    \caption{Maximum accuracy for different number of local rounds ($J$) across combinations under extreme (left column) and moderate (right column) data heterogeneity levels.}
    \label{fig:local-lr-ablation}
\end{figure}

We further explore the effect of changes in the number of local SGD iterations by performing experiments with $J\in\{10, 20\}$. We consider the number of clients to be $N=10$ and test for all data heterogeneity scenarios. The best results for various learning rates $\eta\in\{0.005, 0.001, 0.0005, 0.0001\}$ are included in Fig.~\ref{fig:local-lr-ablation}.

It can be seen from Fig.~\ref{fig:local-lr-ablation} that our method maintains the best performance in all scenarios.
Importantly, the second-best modality under each number of local rounds is different (``I/Q + Phasor" combination when $J=10$ and ``I/Q" modality when $J=20$). This shows that our method, compared to its top rival in each scenario, maintains a robust performance as the number of local rounds changes. Moreover, the results of $\alpha=0.1$ indicate that increasing the number of local SGD iterations does not necessarily always translate into improved performance: in $4$ out of $7$ modality combinations, $J=20$ underperforms 
$J=10$. This behavior represents the effect of increased local model bias induced by excessive local optimization, which becomes more pronounced  under extreme data heterogeneity ($\alpha=0.1$) and leads to a lower performing model after global aggregation.


\subsection{Extension to the ORACLE Dataset} \label{sec:oracle-ablation}
\begin{figure*}[ht]
    \centering
    \includegraphics[width=\textwidth, trim = 0 0 0 17, clip]{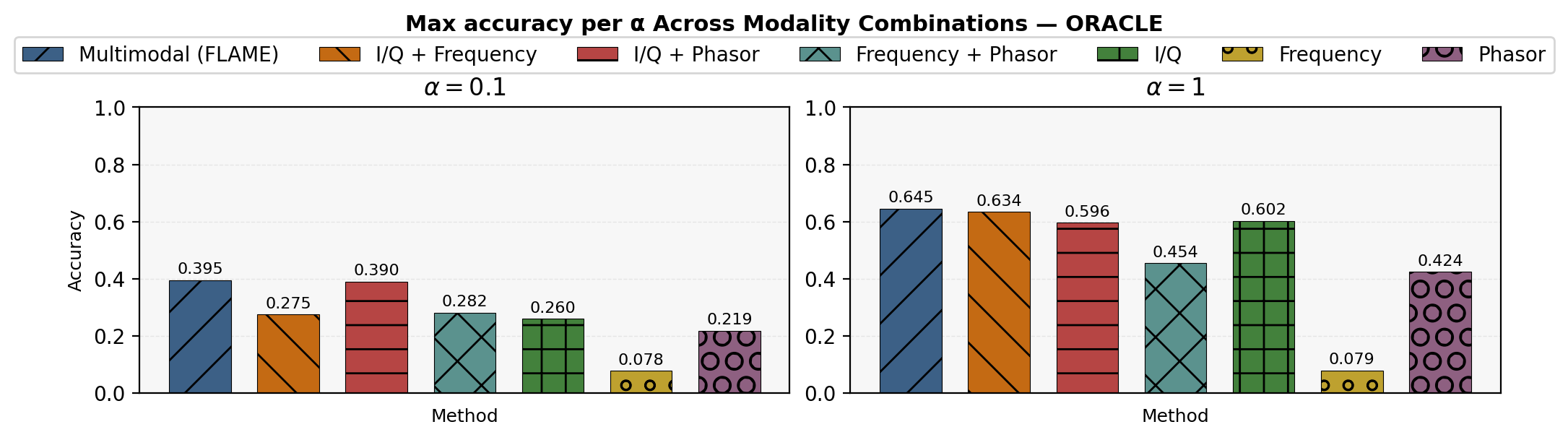}
    \caption{
   Accuracy performance for all modality combinations, across different levels of data heterogeneity on the ORACLE dataset.
    }
    \label{fig:oracle-combination-ablation}
\end{figure*}

While the results on the ORBIT dataset showed the superiority of our multi-modal approach in the accuracy achieved, we expand our experiments to the ORACLE dataset to verify the consistency of performance across datasets. We perform the same experiments as Sec.~\ref{sec:cross-combination} on the ORACLE dataset, with $N=10$, $J=10$, $\eta=0.0001$, and for all modality combinations and data heterogeneity conditions. The results of our experiments are included in Fig.~\ref{fig:oracle-combination-ablation}.

As shown in Fig.~\ref{fig:oracle-combination-ablation}, our method, consistent with the observation in Sec.~\ref{sec:cross-combination}, outperforms the other modality combinations in all data heterogeneity conditions. One key aspect to consider in this plot is how the frequency modality does not seem to be a suitable input for training the network in this case as its accuracy does not improve over different levels of data heterogeneity. This observation is contrary to the findings in Sec.~\ref{sec:cross-combination} which showed improvement in the performance of the frequency modality when trained standalone. We attribute this behavior to the characteristics of the ORACLE dataset, where the unique RF fingerprint may not be captured in the frequency domain.

\begin{figure}[b!]
    \centering
    \includegraphics[width=\linewidth, trim = 0 0 0 19, clip]{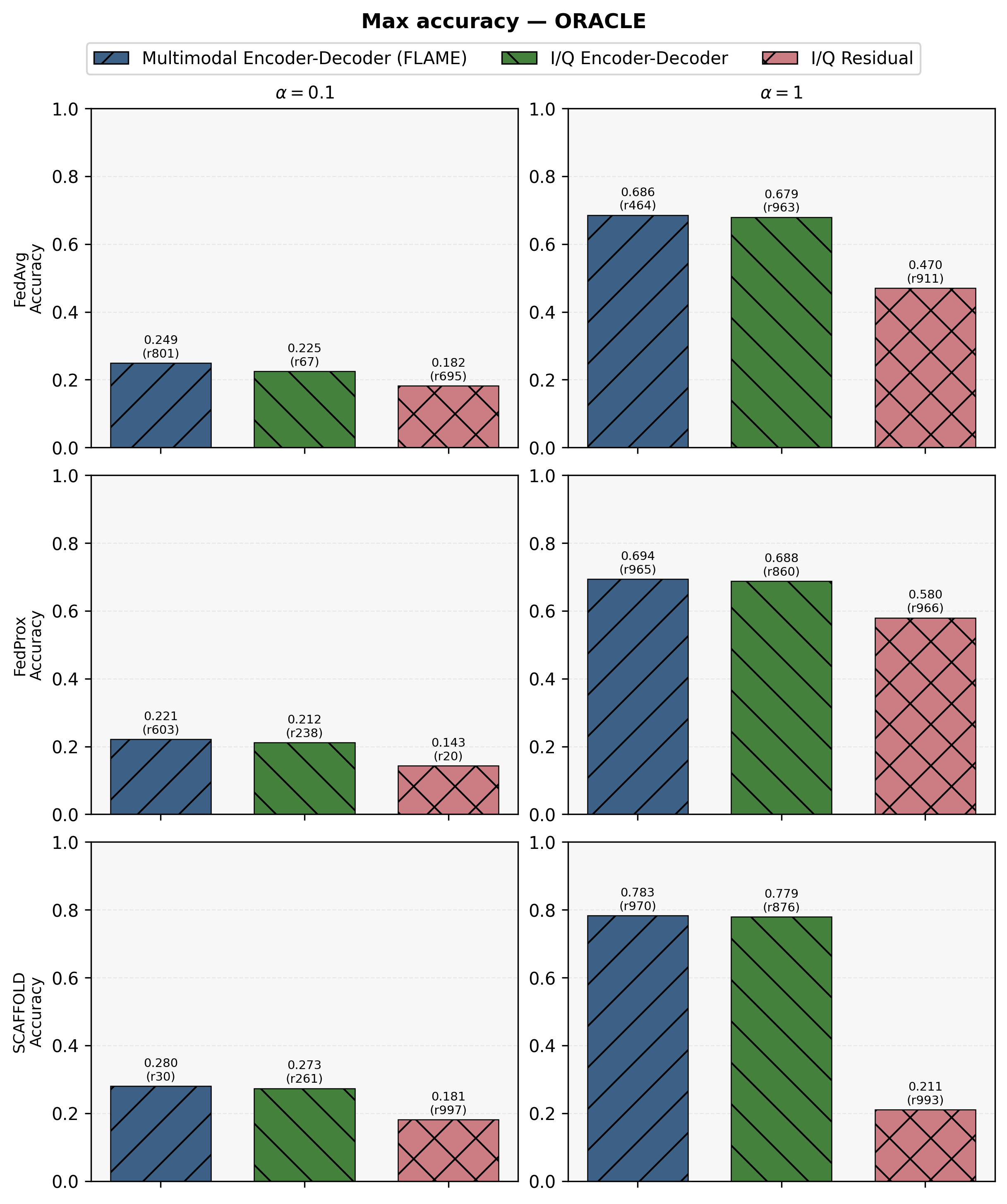}
    \caption{
    Performance of each method across different levels of data heterogeneity for the ORACLE dataset.
    }
    \label{fig:oracle-cross-method}
\end{figure}

We also repeated the comparison across baselines in Sec.~\ref{sec:baselines} for the ORACLE dataset to assess whether the performance gains of our method extend beyond the mere use of multiple modalities. The results of these experiments are included in Fig.~\ref{fig:oracle-cross-method}. Similar to the observations in Sec. IV-C, the objective of this comparison is to verify that the performance improvements introduced by the multi-modal framework remain consistent under each FL method.
Fig. 10 demonstrates that our method consistently outperforms the corresponding single-modal baselines under each FL algorithm (i.e., FedAvg, FedProx, and SCAFFOLD) in both extreme and moderate data heterogeneity settings (Fig. 10-left and -right). While the absolute performance gap may vary depending on the dataset and modality characteristics, the consistent improvement across all FL methods confirms that the gains are attributable to the additional signal representations introduced by our multi-modal design rather than to the choice of FL optimizer.

\subsection{Analyzing the Capabilities of Our Method Through t-SNE}
\label{sec:tsne}
\begin{figure*}[ht]
    \centering
    \includegraphics[width=\textwidth]{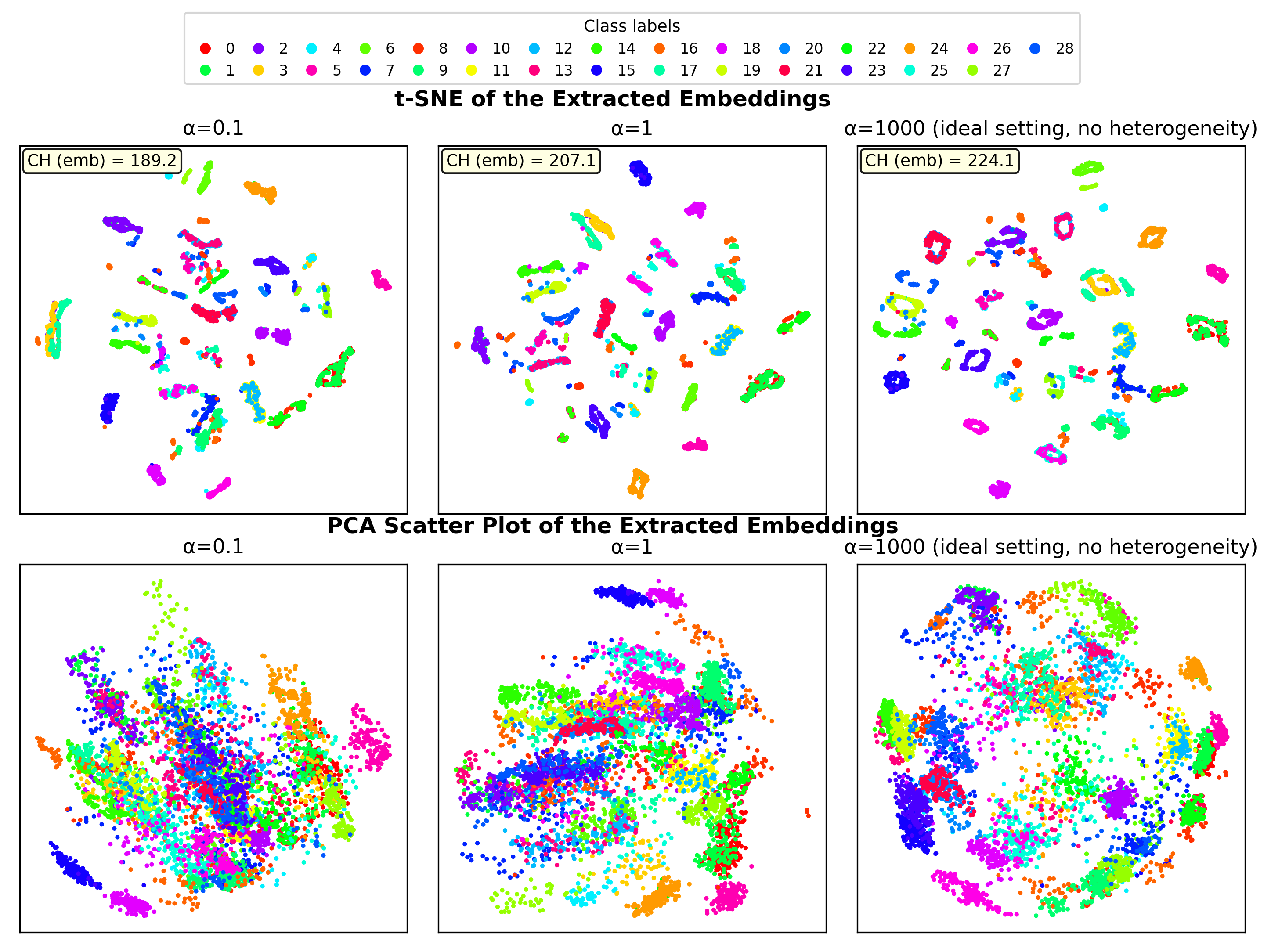}
    \caption{Top Row: t-SNE plots for the extracted embeddings of the encoder-decoder ML model trained for 1000 global rounds under varying levels of data heterogeneity. Datapoints are colored based on their classes. ``CH (emb)" shows the Calinski-Harabasz score which indicates the separation of different classes' embeddings (higher is better). Bottom Row: PCA scatter plot of each class' datapoints' embeddings, colored based on classes.}
    \label{fig:embedding-tsnes}
\end{figure*}

We perform additional experiments to examine how the model’s representation of multi-modal data evolves under different levels of data heterogeneity. To do so, we extract the fused feature embeddings, acquired via the representations computed after modality fusion and before the classification head, from models trained for $1000$ global rounds at data heterogeneity levels $\alpha \in \{0.1, 1, 1000\}$, where we have intentionally plotted the results for an ideal setting with no data heterogeneity ($\alpha=1000$) to better reveal the performance trends. We then visualize these embeddings using t-distributed Stochastic Neighbor Embedding (t-SNE) to assess how samples from different classes cluster in the latent space. To quantify the quality of the clusters formed by the embeddings of each class, we compute the Calinski–Harabasz (CH) score~\cite{calinski1974dendrite}, which measures cluster compactness and separation across classes. For additional insight into how inter-class separation evolves with $\alpha$, we also generate 2D principal component analysis (PCA) scatter plots in which each point is colored according to its ground-truth class. Our results for these experiments are presented in Fig.~\ref{fig:embedding-tsnes}.

As shown, the CH score increases with $\alpha$, indicating that lower data heterogeneity (larger $\alpha$ values) leads to more distinct and well-separated class embeddings. This trend is also visible in the PCA plots in the bottom row of Fig. \ref{fig:embedding-tsnes}: when $\alpha = 0.1$, the embeddings from different classes occupy overlapping regions, yielding less distinguishable clusters. In contrast, when $\alpha = 1000$, the embeddings exhibit clearer spatial separation and more coherent class-specific grouping. Together, these results demonstrate that the global model trained using our proposed method is able to discriminate between classes not only at the final output layer (as reflected in the classification accuracies shown in Figs. \ref{fig:cross-combination-bp} and \ref{fig:cross-combination-ac}) but also within its intermediate fused-representation space, as evidenced by both the CH scores and the embedding visualizations in Fig.~\ref{fig:embedding-tsnes}.

\subsection{Capacity-Matched Single-Modality Baseline} \label{sec:capacity-match}

To isolate the effect of multi-modal information from model capacity, we perform an additional set of capacity-matched single-modality experiments. In Sec.~IV-B, each modality is processed by a dedicated encoder containing a single residual block, resulting in three modality-specific encoders in total. To construct a capacity-matched baseline, we increase the number of residual blocks in the single-modal I/Q encoder to three, matching the total encoder capacity of the multi-modal model.

\begin{figure}[!h]
    \centering
    \includegraphics[width=\linewidth]{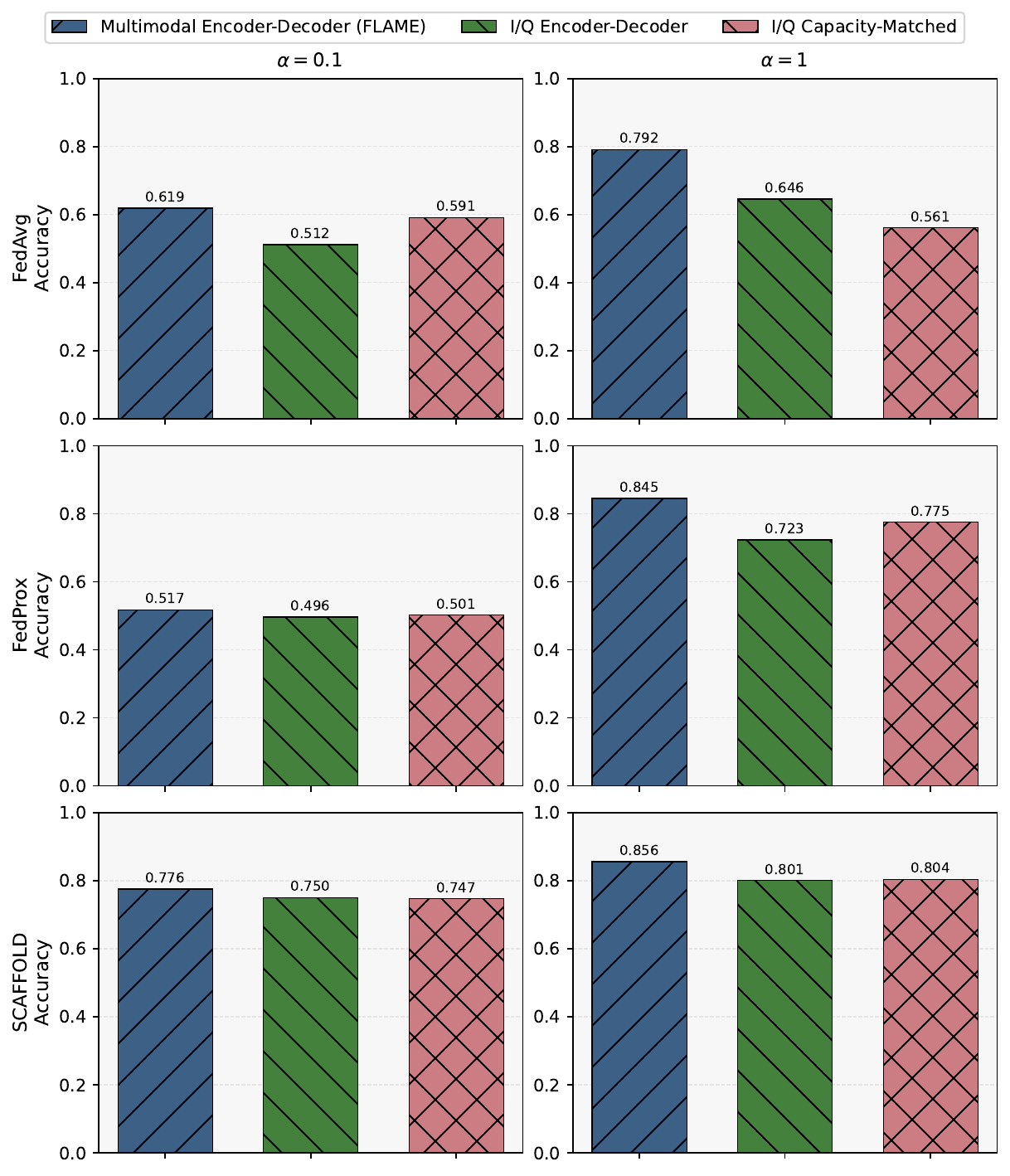}
    \caption{Comparison of the performance of FLAME versus I/Q modality under various encoder capacities.}
    \label{fig:capacity-matched}
\end{figure}

We evaluate this capacity-matched single-modal baseline under the same experimental protocol used for FLAME. Specifically, we repeat the experiments across multiple FL algorithms, including FedAvg, FedProx, and SCAFFOLD, while keeping all datasets, hyperparameters, and training settings identical to those described in Sec.~IV-A. The results are shown in Fig.~\ref{fig:capacity-matched}.
As shown in Fig.~\ref{fig:capacity-matched}, the multi-modal FLAME framework consistently outperforms the capacity-matched single-modal baseline across all FL methods. For example, under FedAvg (Fig.~\ref{fig:capacity-matched} - top), the multi-modal model achieves $61.9\%$ accuracy, whereas the capacity-matched single-modal model reaches $59.1\%$. The same trend holds for FedProx and SCAFFOLD. Note that the accuracy drop in the top-right plot of Fig.~\ref{fig:capacity-matched}, when comparing the capacity-matched model ($56.1\%$) with the lower-capacity single-modal model ($64.6\%$), is likely due to overfitting, which arises from the excessive number of tunable parameters in the model.

These results indicate that the gains achieved by FLAME cannot be explained solely by increased model capacity. Rather, they demonstrate that the complementary information provided by multiple RF signal representations plays a key role in improving federated RF fingerprinting performance.

\subsection{Calculating the Communication and Computation Overheads} \label{sec:comm-comp-overheads}
We next study the resource utilization of FLAME compared to the single-modal counterparts with I/Q modality representing the current state-of-the-art in federated RF fingerprinting, with a focus on the most prevalent FL method, i.e., FedAvg. We do so through analyzing the resource efficiency regarding both computation cycles during local training rounds and the number of communication rounds that are needed to reach a specific accuracy. We also perform this study on FLAME under the ORACLE dataset as an auxiliary set of experiments. We discuss our findings in the following.
\subsubsection{Computation Overhead} \label{sec:computation-overhead}
\begin{figure*}[ht]
    \centering
    \includegraphics[width=\textwidth, trim = 0 0 0 31, clip]{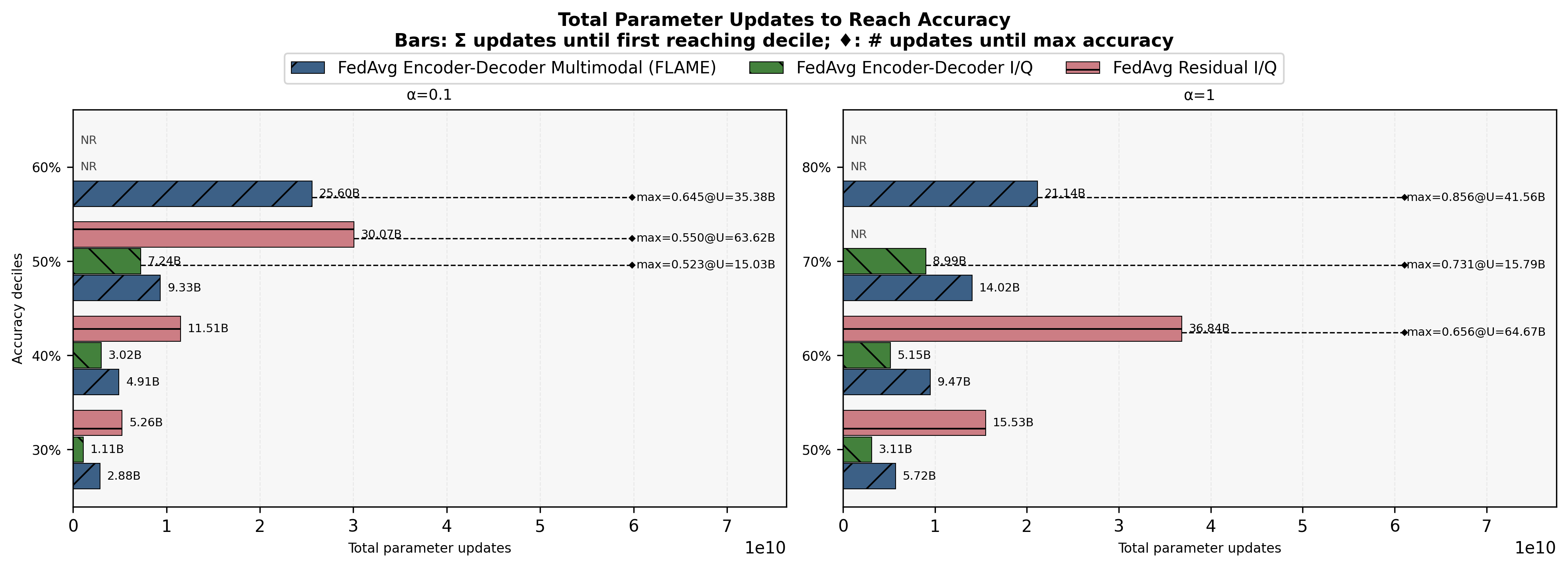}
    \caption{Number of parameter updates for each method to reach an accuracy decile (lower is better), alongside the maximum accuracy reached under ORBIT dataset. Each method (i.e., bar color) is accompanied by its maximum accuracy in the corresponding decile (e.g., ``max=0.645$@$U35.38B" means the maximum accuracy of $64.5\%$ is reached after $35.38$ billion parameters were updated across the network). In each decile, methods that never reached the corresponding accuracy are shown with ``NR" (not reached).}
    \label{fig:comp-overhead}
\end{figure*}

\begin{figure*}[!h]
    \centering
    \includegraphics[width=\textwidth, trim = 0 0 0 31, clip]{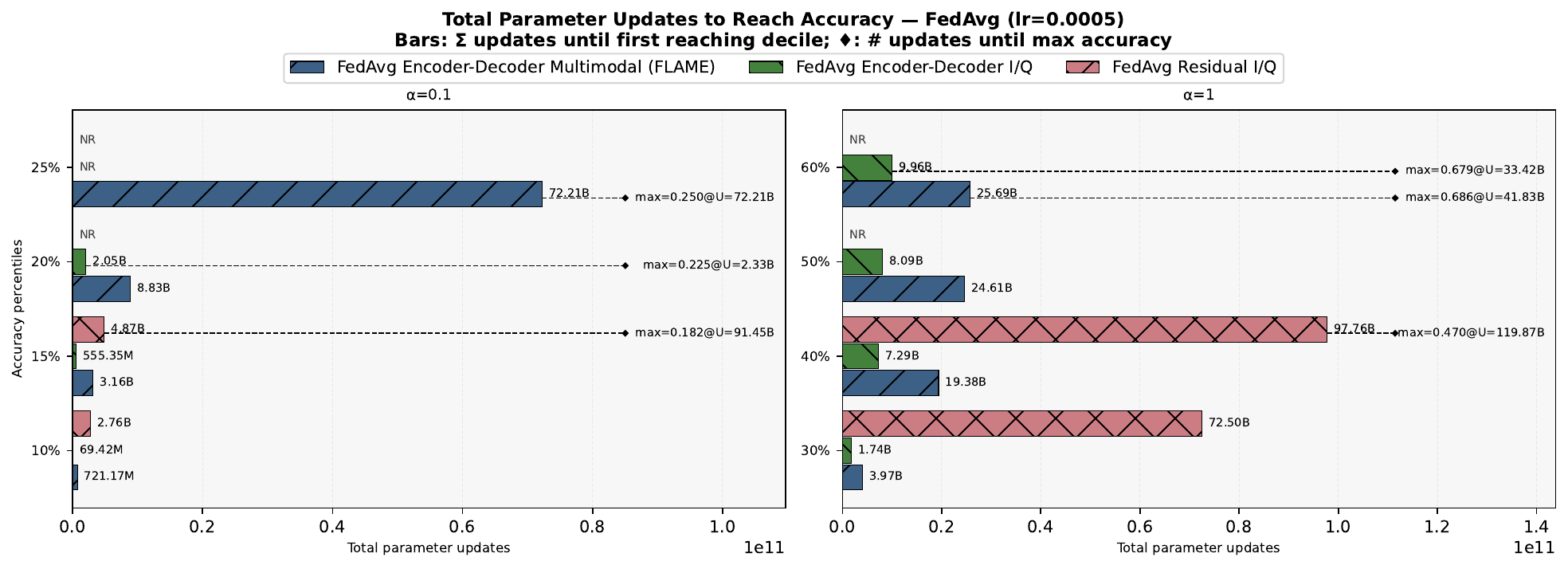}
    \caption{Number of parameter updates for each method to reach an accuracy percentile (lower is better), alongside the maximum accuracy reached under ORACLE dataset. Each method (i.e., bar color) is accompanied by its maximum accuracy in the corresponding percentile (e.g., ``max=0.250$@$U72.21B" means the maximum accuracy of $25.0\%$ is reached after $72.21$ billion parameters were updated across the network). In each percentile, methods that never reached the corresponding accuracy are shown with ``NR" (not reached).}
    \label{fig:comp-overhead-oracle}
\end{figure*}

To study the computation intensity of our method compared to the other scenarios, we plot the number of computations performed for each method, based on the number of updated parameters across the network in the local training stage of each aggregation  round in Fig.~\ref{fig:comp-overhead}. This analysis allows us to examine the relationship between local computation and the performance achieved by each method. Under extreme and moderate data heterogeneity (Fig.~\ref{fig:comp-overhead}-left and -right),
we can see that although encoder–based FedAvg with I/Q modalities requires fewer computations per round, it does not reach the highest performance levels achieved by our multi-modal framework (denoted as “NR” for “Not Reached” in the plots). Performing the same experiments on the ORACLE dataset (Fig.~\ref{fig:comp-overhead-oracle}) resulted in similar results as the ORBIT dataset. Even though the encoder-based I/Q modality combination converges to each performance percentile faster, FLAME outperforms the other baselines in maximum accuracy achieved. This phenomenon is most pronounced under  extreme data heterogeneity (Fig.~\ref{fig:comp-overhead-oracle}-left) where none of the other methods reach the same highest $5\%$ bin (i.e., the 25\% bin). This highlights the trade-off between local computational cost and achievable performance. While FLAME introduces additional local computation due to the use of multiple modality-specific encoders, this increased computational cost enables the model to achieve higher accuracy levels that the lighter baselines fail to attain.

\subsubsection{Communication Overhead} \label{sec:communication-overhead}
\begin{figure*}[ht]
    \centering
    \includegraphics[width=\textwidth, trim = 0 0 0 31, clip]{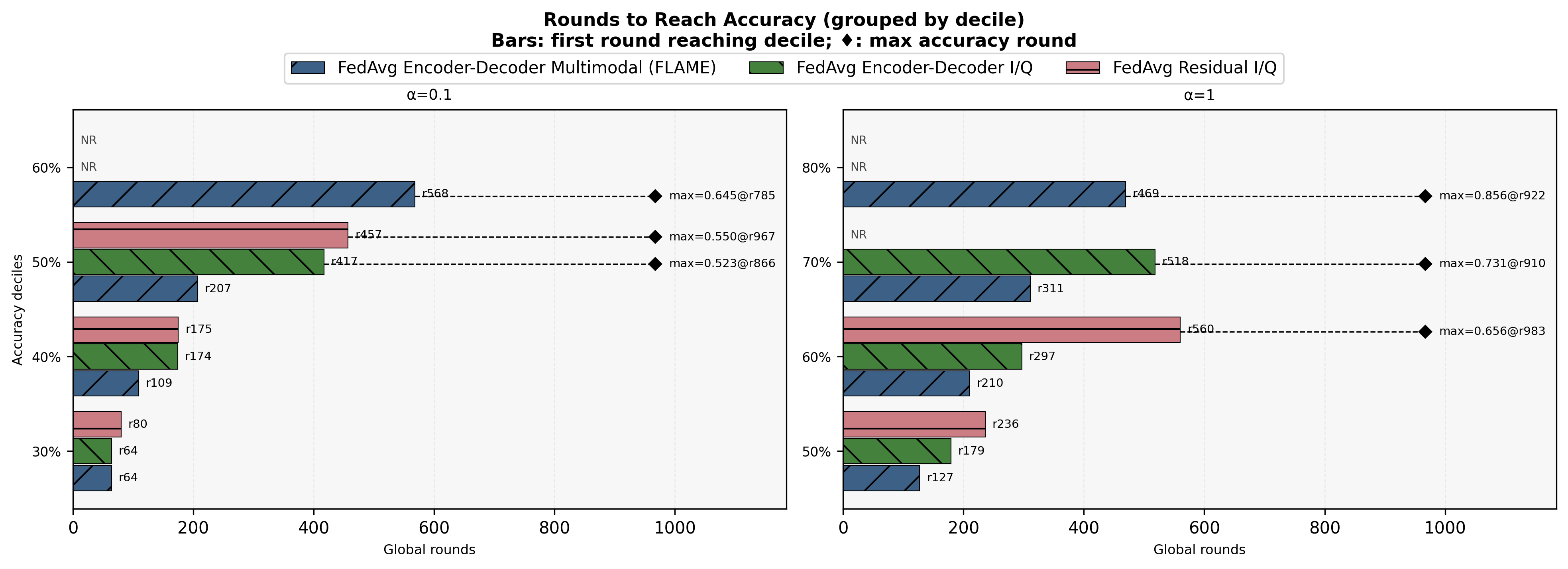}
    \caption{Number of communication rounds for each method to reach an accuracy decile (lower is better), alongside the maximum accuracy reached under ORBIT dataset. Each method (i.e., bar color) is accompanied by its maximum accuracy in the corresponding decile (e.g., ``max=0.645$@$r785" means the maximum accuracy of $64.5\%$ is reached at global round $785$). In each decile, methods that never reached the corresponding accuracy are shown with ``NR" (not reached).}
    \label{fig:comm-overhead}
\end{figure*}

\begin{figure*}[ht]
    \centering
    \includegraphics[width=\textwidth, trim = 0 0 0 31, clip]{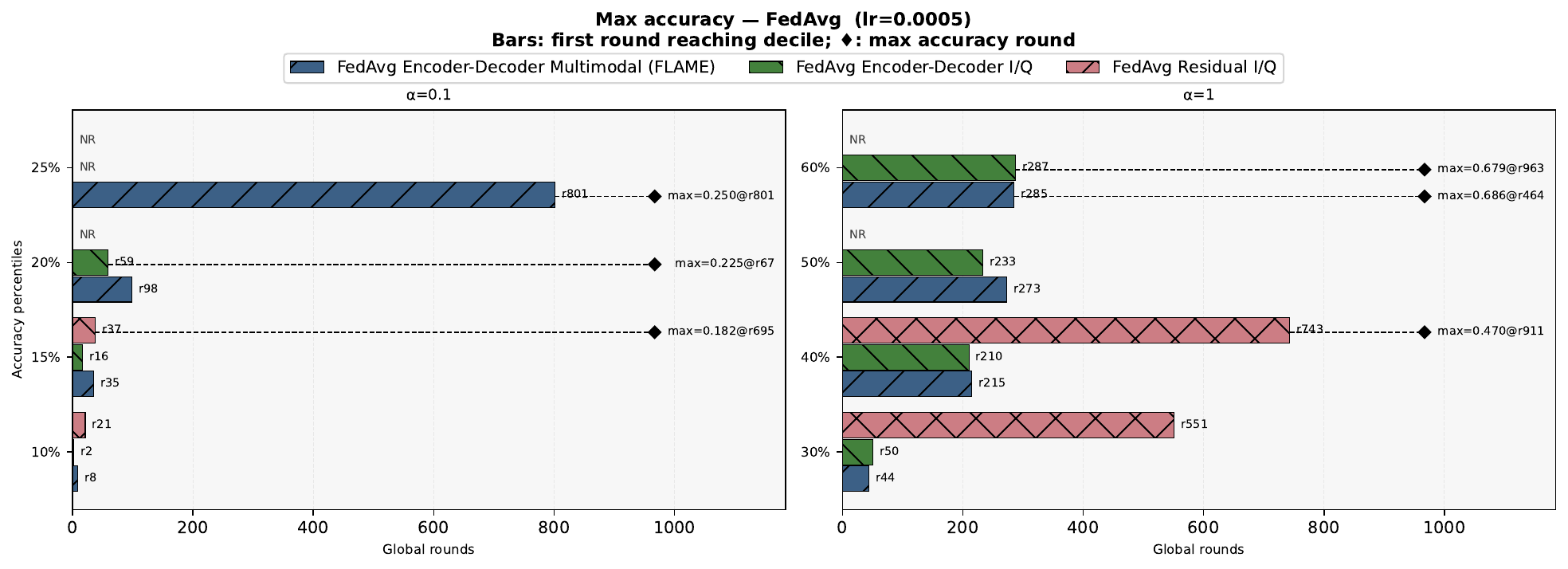}
    \caption{Number of communication rounds for each method to reach an accuracy percentile (lower is better), alongside the maximum accuracy reached under ORACLE dataset. Each method (i.e., bar color) is accompanied by its maximum accuracy in the corresponding percentile (e.g., ``max=0.250$@$r801" means the maximum accuracy of $25.0\%$ is reached at global round $801$). In each percentile, methods that never reached the corresponding accuracy are shown with ``NR" (not reached).}
    \label{fig:comm-overhead-oracle}
\end{figure*}

We perform a study on the number of uplink and downlink communications performed by our method to reach each specific accuracy level. Our results for this comparison are shown in Fig.~\ref{fig:comm-overhead}.
Inspecting Fig.~\ref{fig:comm-overhead}, the number of communication rounds for our method to reach a specific accuracy is shown to be lower than the other methods under both extreme and moderate data heterogeneity, while also reaching the highest accuracy. Repeating the same study on FLAME under the ORACLE dataset (Fig.~\ref{fig:comm-overhead-oracle}), we can see that FLAME reaches the same performance percentiles as encoder-based I/Q modality in approximately the same number of communication rounds. However, the maximum accuracy reached in FLAME stays higher than other methods.

\section{Limitations and Future Work} \label{sec:limitations}

Up to this point, we have been able to show how our proposed method shows a superior performance when compared to the other baselines performing the same function. However, our method in its current form has some limitations and bottlenecks which we acknowledge in this section, and suggest as future directions of research. Notable future directions are:

\begin{itemize}
    \item \textbf{Reducing Computation Overhead}: The gap shown in Sec.~\ref{sec:computation-overhead} motivates future research on computationally efficient methods to maintain the performance level in more extreme data heterogeneity scenarios while reducing the number of computations needed to reach higher accuracies when the data heterogeneity is not severe.
    Other prominent future directions include the investigation of approaches such as Federated Dropout~\cite{xie2024federated} and heterogeneous numbers of local iterations~\cite{narmadha2025fedeff} to reduce the overall computation overhead.
    
    \item \textbf{Reducing Communication Overhead}: 
    Following the results in Sec.~\ref{sec:communication-overhead}, in real-world network settings, bandwidth limitations and transmission power consumption limitations can impact each device's ability to perform uplink and downlink communications. Consequently, further research can be conducted on communication-efficient extensions of our method using techniques such as asynchronous client participation and model aggregation, communication compression~\cite{pmlr-v238-zakerinia24a} and communication overhead regularization~\cite{li2024fedsparse}.
    
    \item \textbf{Robust Multi-Modal FL}:  Signal perturbations arising from channel noise or adversarial inference can corrupt local data and, in turn, poison the federated training process. These robustness challenges are largely orthogonal to the focus of this work, which centers on extending FL from single-modal to multi-modal settings for RF fingerprinting. Nevertheless, extending our framework to scenarios where multi-modal data are subject to heterogeneous channel impairments and adversarial inference attacks is an important and promising direction for future research. Furthermore, although FL preserves data locality, adversarial methods such as data reconstruction attacks can still extract sensitive information from the gradients exchanged during model aggregation. Accordingly, further research is needed to develop and integrate privacy-preserving defenses (e.g., differential privacy or secure aggregation) into multi-modal FL frameworks for RF fingerprinting.
    \item \textbf{Model Heterogeneity}: The computation/storage heterogeneity across devices can in turn impose heterogeneity in their model training resources. In practical deployments, such resource constraints may also limit the number of modalities that a client can compute, resulting in scenarios where a client only has access to a subset of modalities (i.e., $M_n < M$). One way to address this issue can be addressed through deploying varying models on dissimilar devices which require power, memory, and computation usage according to the device they are deployed on. However, such model heterogeneity fundamentally impacts core operations such as model aggregation and must therefore be carefully addressed: for example, partial model sharing strategies could be employed where only selected components of the model (e.g., modality-specific encoders corresponding to the available modalities) are trained and aggregated across devices. Additional mechanisms such as modality-aware aggregation or missing-modality handling strategies may also be required to maintain consistent global representations. Future research can thus focus on developing such methods to increase the deployability of multi-modal FL for RF fingerprinting on resource-constrained devices.
    \item \textbf{Improved Fusion Techniques}: In this work, we have considered passing the concatenated features extracted from the modality-specific encoders through a fully-connected layer as our fusion method. However, depending on factors such as the dataset, modality composition, transmitter type, and modulation scheme, different modalities may contribute unequally to the learning process and exhibit varying levels of influence on model performance. Moreover, within each modality, different feature components extracted by the encoders may carry differing degrees of relevance for the final prediction. To better exploit these relationships and further improve performance, future work may investigate more expressive fusion mechanisms, such as attention-based fusion.
    \item \textbf{Partial Modalities}: Given the possible resource constraints and signal quality variations across the network, some devices may not be able to calculate the representations for the frequency or phasor modalities or both. To alleviate this barrier, future research can focus on adapting multi-modal FL in the presence of clients with partial/missing modalities for RF fingerprinting.
    
\end{itemize}

\section{Conclusion} \label{conclusion_sec}
In this paper, we proposed FLAME: a federated learning (FL) approach for multi-modal RF fingerprinting. We showed that our proposed framework increases the convergence rate and overall global classification performance in a variety of federated learning settings when applied to fingerprinting data at multiple wireless access points (APs). We theoretically determined a convergence bound and empirically validated its efficacy on a real-world dataset. We then performed personalized federated learning and fine-tuned the local model at each AP after the conclusion of the FL training rounds to maximize the fingerprinting classification performance on the distribution of data at each AP. In this setting, we showed that each AP benefits from a higher local accuracy after personalized federated learning is applied, thus making it a vital component of the FLAME framework. In future work, we anticipate extending our framework to consider its effectiveness on smaller and larger sets of transmitters. We also anticipate exploring the behavior of our derived convergence bound on out-of-distribution waveforms such as varying channel conditions and adversarial examples. 


\bibliography{references}

\bibliographystyle{IEEEtran}

\appendices

\section{Proof of Lemma~\ref{lemma:11}}
For notational simplicity, define
$
\mathbf{u}_{m}
\triangleq
\mathbf{g}_{n,m}^{\mathrm{Sh}}(\mathbf{w}),
\qquad
A
\triangleq
\sum_{m=1}^{M}
\|\mathbf{u}_{m}\|^2.
$
First, since
$
2\langle \mathbf{u}_{m},\mathbf{u}_{m'}\rangle
\leq
\|\mathbf{u}_{m}\|^2+\|\mathbf{u}_{m'}\|^2,
$
summing over all pairs $m<m'$ yields
$
2
\sum_{m=1}^{M-1} \sum_{m'=m+1}^{M}
\langle \mathbf{u}_{m},\mathbf{u}_{m'}\rangle
\leq
(M-1)
\sum_{m=1}^{M}
\|\mathbf{u}_{m}\|^2
=
(M-1)A.
$
Therefore,
\begin{equation}
\varrho_n(\mathbf{w})
\leq
\frac{(M-1)A}{(M-1)A+\epsilon}
<1.
\end{equation}

Next, using the identity
$
\left\|
\sum_{m=1}^{M}
\mathbf{u}_{m}
\right\|^2
=
\sum_{m=1}^{M}
\|\mathbf{u}_{m}\|^2
+
2
\sum_{m=1}^{M-1} \sum_{m'=m+1}^{M}
\langle \mathbf{u}_{m},\mathbf{u}_{m'}\rangle,
$
and the fact that squared norms are nonnegative, we obtain
$
A
+
2
\sum_{m=1}^{M-1} \sum_{m'=m+1}^{M}
\langle \mathbf{u}_{m},\mathbf{u}_{m'}\rangle
\geq
0.
$
Hence,
$
2
\sum_{m=1}^{M-1} \sum_{m'=m+1}^{M}
\langle \mathbf{u}_{m},\mathbf{u}_{m'}\rangle
\geq
-A.
$
It follows that
$
\varrho_n(\mathbf{w})
\geq
\frac{-A}{(M-1)A+\epsilon}
>
-\frac{1}{M-1}.
$
Combining the aforementioned upper and lower bounds gives
\begin{equation}
    -\frac{1}{M-1}
<
\varrho_n(\mathbf{w})
<
1.
\end{equation}
Since $M\geq 2$, we also have $-\frac{1}{M-1}\geq -1$, and therefore
$
-1
<
\varrho_n(\mathbf{w})
<
1.
$
This completes the proof.

\section{Proof of Theorem~\ref{thm2}}
\label{subsec:mm_convergence_ingredients}

 Using the $L$-Lipschitz smoothness of $f(\mathbf{w})$, we have
\begin{equation} 
    f(\mathbf{w}^{(t+1)}) \leq f(\mathbf{w}^{(t)}) \nonumber
\end{equation}
\begin{equation} \label{lip}
    + \langle \nabla f(\mathbf{w}^{(t)}), \mathbf{w}^{(t+1)}-\mathbf{w}^{(t)}\rangle + \frac{L}{2} ||\mathbf{w}^{(t+1)}-\mathbf{w}^{(t)}||^2. 
\end{equation}
Given that local parameter updates are given by (\ref{local_update}), the final local model after $J$ SGD steps is given by
\begin{equation} \label{local_update_ex}
    \mathbf{w}^{(t, J)}_{n} = \mathbf{w}^{(t)} - \eta \sum_{j=0}^{J-1}\widetilde{\nabla} f_{n}(\mathbf{w}^{(t, j)}_{n}). 
\end{equation}
Substituting (\ref{local_update_ex}) into the parameter update expression in (\ref{agg_eqn}), we see that 
\begin{equation}
    \mathbf{w}^{(t+1)} = \frac{1}{N}\sum_{n=1}^{N} \bigg{(} \mathbf{w}^{(t)} - \eta \sum_{j=0}^{J-1}\widetilde{\nabla} f_{n}(\mathbf{w}^{(t, j)}_{n})\bigg{)} \nonumber 
\end{equation}
\begin{equation}
    \quad \hspace{2mm} = \mathbf{w}^{(t)} - \frac{\eta}{N} \sum_{n=1}^{N} \sum_{j=0}^{J-1} \widetilde{\nabla} f_{n}(\mathbf{w}^{(t, j)}_{n}),
\end{equation}
and thus
\begin{equation} \label{w_diff}
    \mathbf{w}^{(t+1)} - \mathbf{w}^{(t)} = -\frac{\eta}{N} \sum_{n=1}^{N} \sum_{j=0}^{J-1} \widetilde{\nabla} f_{n}(\mathbf{w}^{(t, j)}_{n}).
\end{equation}
Substituting (\ref{w_diff}) into the smoothness inequality from (\ref{lip}), and by taking the expectation of both hand sides (with respect to the SGD sampling at aggregation round $t$; note that $\mathbb{E}_t[\widetilde{\nabla} f_{n}(\mathbf{w}^{(t, j)}_{n})]={\nabla} f_{n}(\mathbf{w}^{(t, j)}_{n})$), we get:
\begin{equation} \label{pin:0}
\begin{aligned}
    &\mathbb{E}_t[f(\mathbf{w}^{(t+1)})] \leq  f(\mathbf{w}^{(t)}) \\
    &
    - \eta \bigg{\langle} \nabla f(\mathbf{w}^{(t)}), \frac{1}{N}\sum_{n=1}^{N} \sum_{j=0}^{J-1} {\nabla} f_{n}(\mathbf{w}^{(t, j)}_{n})\bigg{\rangle} \\
    &
    + \frac{L}{2} \mathbb{E}_t\Bigg[\bigg{|}\bigg{|} -\frac{\eta}{N} \sum_{n=1}^{N} \sum_{j=0}^{J-1} \widetilde{\nabla} f_{n}(\mathbf{w}^{(t, j)}_{n}) \bigg{|}\bigg{|}^2\Bigg].
    \end{aligned}
\end{equation}
Note that 
\begin{equation}
\hspace{-3mm}
\resizebox{0.89\linewidth}{!}{$
    \begin{aligned}
        &\mathbb{E}_t\Bigg[\bigg{|}\bigg{|} -\frac{\eta}{N} \sum_{n=1}^{N} \sum_{j=0}^{J-1} \widetilde{\nabla} f_{n}(\mathbf{w}^{(t, j)}_{n}) \bigg{|}\bigg{|}^2\Bigg]=
        \\&\eta^2\mathbb{E}_t\Bigg[\bigg{|}\bigg{|} -\frac{1}{N} \sum_{n=1}^{N} \sum_{j=0}^{J-1} \widetilde{\nabla} f_{n}(\mathbf{w}^{(t, j)}_{n})+\frac{1}{N} \sum_{n=1}^{N} \sum_{j=0}^{J-1} {\nabla} f_{n}(\mathbf{w}^{(t, j)}_{n})\\
        &-\frac{1}{N} \sum_{n=1}^{N} \sum_{j=0}^{J-1} {\nabla} f_{n}(\mathbf{w}^{(t, j)}_{n}) \bigg{|}\bigg{|}^2\Bigg]\\
        &=\eta^2 \mathbb{E}_t\Bigg[\bigg{|}\bigg{|}\frac{1}{N} \sum_{n=1}^{N} \sum_{j=0}^{J-1}{\nabla} f_{n}(\mathbf{w}^{(t, j)}_{n}) \bigg{|}\bigg{|}^2\Bigg]+J \eta^2\sigma_{\mathrm{MM}}^2,
    \end{aligned} 
    $}
\end{equation}
where in the last equality we have used the zero-mean property of SGD noise. 
Replacing the above result in~\eqref{pin:0} yields: 
\begin{equation} \label{pin:01}
\hspace{-3mm}
\resizebox{0.89\linewidth}{!}{$
\begin{aligned}
    &\mathbb{E}_t[f(\mathbf{w}^{(t+1)})] \leq  f(\mathbf{w}^{(t)})\\
    &
     \underbrace{- \eta J \bigg{\langle} \nabla f(\mathbf{w}^{(t)}), \frac{1}{NJ}\sum_{n=1}^{N} \sum_{j=0}^{J-1} {\nabla} f_{n}(\mathbf{w}^{(t, j)}_{n})\bigg{\rangle}}_{(a)} 
     \\&
    + \frac{L\eta^2J^2}{2}  \mathbb{E}_t\Bigg[\bigg{|}\bigg{|} \frac{1}{NJ} \sum_{n=1}^{N} \sum_{j=0}^{J-1} {\nabla} f_{n}(\mathbf{w}^{(t, j)}_{n}) \bigg{|}\bigg{|}^2\Bigg]+\frac{L}{2}J\eta^2\sigma_{\mathrm{MM}}^2,
    \end{aligned}
    $}
\end{equation}

To bound term $(a)$, we use the fact that $\Vert A-B\Vert^2=\Vert A \Vert^2 + \Vert B \Vert^2 -2 \langle A,B\rangle \Rightarrow \langle A,B\rangle= \frac{1}{2}\Vert A \Vert^2 + \frac{1}{2}\Vert B \Vert^2 - \frac{1}{2} \Vert A-B\Vert^2$
and we get:
\begin{equation}
\hspace{-3mm}
\resizebox{0.89\linewidth}{!}{$
\begin{aligned}
    (a)=&-\eta J\frac{1}{2} \left\Vert \nabla f(\mathbf{w}^{(t)})\right\Vert^2-\eta J\frac{1}{2}\left\Vert \frac{1}{NJ}\sum_{n=1}^{N} \sum_{j=0}^{J-1} {\nabla} f_{n}(\mathbf{w}^{(t, j)}_{n}) \right\Vert^2 \\&+ \eta J\frac{1}{2} \left\Vert \nabla f(\mathbf{w}^{(t)}) - \frac{1}{NJ}\sum_{n=1}^{N} \sum_{j=0}^{J-1} {\nabla} f_{n}(\mathbf{w}^{(t, j)}_{n}) \right\Vert^2.
    \end{aligned}
    $}
\end{equation}
Replacing the above result in~\eqref{pin:01} and factoring out the common terms yield:
\begin{equation}
\begin{aligned}
    &f(\mathbf{w}^{(t+1)}) \leq  f(\mathbf{w}^{(t)})\\
&-\eta J \frac{1}{2} \left\Vert \nabla f(\mathbf{w}^{(t)})\right\Vert^2 +\frac{L}{2}J\eta^2\sigma_{\mathrm{MM}}^2\\&+ 
\underbrace{\left(-\frac{\eta J}{2}+\frac{L\eta^2 J^2}{2}\right)}_{(b)}
\left\Vert \frac{1}{NJ} \sum_{n=1}^{N} \sum_{j=0}^{J-1} {\nabla} f_{n}(\mathbf{w}^{(t, j)}_{n}) \right\Vert^2
\\
&+\eta J\frac{1}{2} \left\Vert \nabla f(\mathbf{w}^{(t)}) - \frac{1}{NJ}\sum_{n=1}^{N} \sum_{j=0}^{J-1} {\nabla} f_{n}(\mathbf{w}^{(t, j)}_{n}) \right\Vert^2.
    \end{aligned}
\end{equation}
Under the choice of step size $\eta<\frac{1}{JL}$, the coefficient $(b)$ becomes negative and thus its corresponding term can be eliminated from the upper-bound. Thus, with taking the expectation from both hand sides (with respect to the SGD sampling at aggregation round $t$), we get:
 \begin{equation}\label{pin:2}
 \hspace{-3mm}
\resizebox{0.89\linewidth}{!}{$
\begin{aligned}
    &\mathbb{E}_t[f(\mathbf{w}^{(t+1)})] \leq  f(\mathbf{w}^{(t)})\\
&-\eta J \frac{1}{2} \left\Vert \nabla f(\mathbf{w}^{(t)})\right\Vert^2+\frac{L}{2}J\eta^2\sigma_{\mathrm{MM}}^2\\
&+\underbrace{\eta J \frac{1}{2} \mathbb{E}_t\left[\left\Vert \nabla f(\mathbf{w}^{(t)}) - \frac{1}{NJ}\sum_{n=1}^{N} \sum_{j=0}^{J-1} {\nabla} f_{n}(\mathbf{w}^{(t, j)}_{n}) \right\Vert^2\right]}_{(c)}.
    \end{aligned}
    $}
\end{equation}
We then bound term $(c)$ as follows:
 \begin{equation}\label{pin:1}
 \hspace{-3mm}
\resizebox{0.89\linewidth}{!}{$
\begin{aligned}
  &\eta J \frac{1}{2} \mathbb{E}_t\left[\left\Vert \nabla f(\mathbf{w}^{(t)}) - \frac{1}{NJ}\sum_{n=1}^{N} \sum_{j=0}^{J-1} {\nabla} f_{n}(\mathbf{w}^{(t, j)}_{n}) \right\Vert^2\right]\\
  &=\eta J\frac{1}{2} \mathbb{E}_t\left[\left\Vert \frac{1}{NJ}\sum_{n=1}^{N}\sum_{j=0}^{J-1}\nabla f_n(\mathbf{w}^{(t)}) - \frac{1}{NJ}\sum_{n=1}^{N} \sum_{j=0}^{J-1} 
  {\nabla} f_{n}(\mathbf{w}^{(t, j)}_{n}) \right\Vert^2\right]
  \\
  & \leq \eta J\frac{1}{2} \frac{1}{NJ}\sum_{n=1}^{N}\sum_{j=0}^{J-1}\mathbb{E}_t\left[\left\Vert \nabla f_n(\mathbf{w}^{(t)}) -   
  {\nabla} f_{n}(\mathbf{w}^{(t, j)}_{n}) \right\Vert^2\right]\\
  &
  =\eta\frac{1}{2} \frac{1}{N}\sum_{n=1}^{N}\sum_{j=0}^{J-1}\mathbb{E}_t\left[\left\Vert \nabla f_n(\mathbf{w}^{(t)}) -  
  {\nabla} f_{n}(\mathbf{w}^{(t, j)}_{n}) \right\Vert^2\right]\\
  &
  \leq \eta \frac{1}{2} \frac{1}{N}L^2\sum_{n=1}^{N}\sum_{j=0}^{J-1} \underbrace{\mathbb{E}_t\left[\left\Vert \mathbf{w}^{(t)} -  
 \mathbf{w}^{(t, j)}_{n} \right\Vert^2\right]}_{(d)}
    \end{aligned}
    $}
\end{equation}
where the first inequality is obtained using Jensen’s inequality and the last inequality is obtained using the $L-$smoothness property. We next bound term $(d)$ as follows:
\begin{equation}
\hspace{-3mm}
\resizebox{0.89\linewidth}{!}{$
    \begin{aligned}
        &\mathbb{E}_t\left[\left\Vert \mathbf{w}^{(t)} -  
 \mathbf{w}^{(t, j)}_{n} \right\Vert^2\right]=\mathbb{E}_t\left[\left\Vert 
  \eta \sum_{j'=0}^{j-1} \widetilde{\nabla} f_{n}(\mathbf{w}^{(t, j')}_{n})
  \right\Vert^2\right]
  \\
  &= \eta^2\mathbb{E}_t\left[\left\Vert 
  \sum_{j'=0}^{j-1} \widetilde{\nabla} f_{n}(\mathbf{w}^{(t, j')}_{n})
  \right\Vert^2\right]\\
  &= \eta^2\mathbb{E}_t\left[\left\Vert 
  \sum_{j'=0}^{j-1} [\widetilde{\nabla} f_{n}(\mathbf{w}^{(t, j')}_{n})-{\nabla} f_{n}(\mathbf{w}^{(t, j')}_{n})+{\nabla} f_{n}(\mathbf{w}^{(t, j')}_{n})]
  \right\Vert^2\right]
  \\
  & \leq 2 \eta^2\mathbb{E}_t\left[\left\Vert 
  \sum_{j'=0}^{j-1} [\widetilde{\nabla} f_{n}(\mathbf{w}^{(t, j')}_{n})-{\nabla} f_{n}(\mathbf{w}^{(t, j')}_{n})]
  \right\Vert^2\right] \\&
  + 2\eta^2 \mathbb{E}_t\left[\left\Vert 
  \sum_{j'=0}^{j-1} {\nabla} f_{n}(\mathbf{w}^{(t, j')}_{n})
  \right\Vert^2\right]
  \\
  & 
  \leq 2 \eta^2 \sum_{j'=0}^{j-1} \sigma_{\mathrm{MM}}^2 
  + 2\eta^2 \mathbb{E}_t\left[\left\Vert 
  \sum_{j'=0}^{j-1} {\nabla} f_{n}(\mathbf{w}^{(t, j')}_{n})
  \right\Vert^2\right]
  \\
  &
  \leq  2 \eta^2 \sum_{j'=0}^{j-1} \sigma_{\mathrm{MM}}^2 
  {+} 2\eta^2 j  \sum_{j'=0}^{j-1} \mathbb{E}_t\left[\left\Vert 
  {\nabla} f_{n}(\mathbf{w}^{(t, j')}_{n}) -{\nabla} f_n(\mathbf{w}^{(t)})+{\nabla} f_n(\mathbf{w}^{(t)})
  \right\Vert^2\right]
  \\&
  \leq  2 \eta^2 \sum_{j'=0}^{j-1} \sigma_{\mathrm{MM}}^2 
  {+} 4\eta^2 j  \sum_{j'=0}^{j-1} \mathbb{E}_t\left[\left\Vert 
  {\nabla} f_{n}(\mathbf{w}^{(t, j')}_{n}) -{\nabla} f_n(\mathbf{w}^{(t)}) \right\Vert^2\right]
  \\&
  +4\eta^2 j  \sum_{j'=0}^{j-1} \mathbb{E}_t
  \left[\left\Vert 
  {\nabla} f_n(\mathbf{w}^{(t)})
  \right\Vert^2\right]
   \\&
  \leq  2 \eta^2 j \sigma_{\mathrm{MM}}^2 
  {+} 4\eta^2 j L^2  \sum_{j'=0}^{j-1} \mathbb{E}_t\left[\left\Vert 
  \mathbf{w}^{(t, j')}_{n} -\mathbf{w}^{(t)} \right\Vert^2\right]
  \\&
  +4\eta^2 j  \sum_{j'=0}^{j-1} \left\Vert 
  {\nabla} f_n(\mathbf{w}^{(t)})
  \right\Vert^2
    \end{aligned}
    $}
\end{equation}
where the inequalities were obtained using Cauchy–Schwarz inequality, $L-$smoothness property, and the SGD noise properties.  The above bound implies:
\begin{equation}
\hspace{-3mm}
\resizebox{0.89\linewidth}{!}{$
    \begin{aligned}
        &\sum_{j=0}^{J-1}\mathbb{E}_t\left[\left\Vert \mathbf{w}^{(t)} -  
 \mathbf{w}^{(t, j)}_{n} \right\Vert^2\right]
  \\& \leq  2 \eta^2 \sum_{j=0}^{J-1} j \sigma_{\mathrm{MM}}^2 
  {+} 4\eta^2L^2 \sum_{j=0}^{J-1} j   \sum_{j'=0}^{j-1} \mathbb{E}_t\left[\left\Vert 
  \mathbf{w}^{(t, j')}_{n} -\mathbf{w}^{(t)} \right\Vert^2\right]
  \\&
  +4\eta^2 \sum_{j=0}^{J-1} j  \sum_{j'=0}^{j-1} \mathbb{E}_t
  \left[\left\Vert 
  {\nabla} f_n(\mathbf{w}^{(t)})
  \right\Vert^2\right]\\
  \\&\leq 2 \eta^2 \frac{J(J-1)}{2} \sigma_{\mathrm{MM}}^2   {+} 4\eta^2L^2 J (J-1)   \sum_{j=0}^{J-1} \mathbb{E}_t\left[\left\Vert 
  \mathbf{w}^{(t, j)}_{n} -\mathbf{w}^{(t)} \right\Vert^2\right]
  \\& +4\eta^2 J^2(J-1) \left\Vert 
  {\nabla} f_n(\mathbf{w}^{(t)})
  \right\Vert^2
  \\
  &\Longrightarrow  \sum_{j=0}^{J-1}\mathbb{E}_t\left[\left\Vert \mathbf{w}^{(t)} -  
 \mathbf{w}^{(t, j)}_{n} \right\Vert^2\right]\\&\leq 2 \eta^2 \frac{J(J-1)}{2} \sigma_{\mathrm{MM}}^2   {+} 4\eta^2L^2 J (J-1)   \sum_{j=0}^{J-1} \mathbb{E}_t\left[\left\Vert 
  \mathbf{w}^{(t, j)}_{n} -\mathbf{w}^{(t)} \right\Vert^2\right]
  \\& +4\eta^2 J^2(J-1) \left\Vert 
  {\nabla} f_n(\mathbf{w}^{(t)})
  \right\Vert^2.
    \end{aligned}
    $}
\end{equation}
Assuming   $\eta \leq \frac{1}{2L\sqrt{J(J-1)}}$, the above result yields:
\begin{equation}
    \begin{aligned}
        &\sum_{j=0}^{J-1}\mathbb{E}_t\left[\left\Vert \mathbf{w}^{(t)} -  
 \mathbf{w}^{(t, j)}_{n} \right\Vert^2\right] 
  \\& \leq \frac{2 \eta^2 \frac{J(J-1)}{2} \sigma_{\mathrm{MM}}^2 }{1-4\eta^2L^2 J (J-1)}
  \\&
  +\frac{4\eta^2 J^2(J-1)}{1-4\eta^2L^2 J (J-1)} \left\Vert 
  {\nabla} f_n(\mathbf{w}^{(t)}) \right\Vert^2.
    \end{aligned}
\end{equation}
Replacing the above result back in~\eqref{pin:1}, we get:
\begin{equation}
\hspace{-3mm}
\resizebox{0.89\linewidth}{!}{$
    \begin{aligned}
        &\eta J \frac{1}{2} \mathbb{E}_t\left[\left\Vert \nabla f(\mathbf{w}^{(t)}) - \frac{1}{NJ}\sum_{n=1}^{N} \sum_{j=0}^{J-1} \widetilde{\nabla} f_{n}(\mathbf{w}^{(t, j)}_{n}) \right\Vert^2\right]\\
        & \leq \eta \frac{1}{2} \frac{1}{N}L^2\sum_{n=1}^{N}
        \frac{2 \eta^2 \frac{J(J-1)}{2} \sigma_{\mathrm{MM}}^2 }{1-4\eta^2L^2 J (J-1)}
        \\&
        + \eta\frac{1}{2}  \frac{1}{N}L^2\sum_{n=1}^{N}\frac{4\eta^2 J^2(J-1)}{1-4\eta^2L^2 J (J-1)} \left\Vert 
  {\nabla} f_n(\mathbf{w}^{(t)}) \right\Vert^2
  \\& =\eta\frac{1}{2}  L^2
        \frac{2 \eta^2 \frac{J(J-1)}{2} \sigma_{\mathrm{MM}}^2 }{1-4\eta^2L^2 J (J-1)}
        \\&
        + \eta\frac{1}{2}L^2\frac{4\eta^2 J^2(J-1)}{1-4\eta^2L^2 J (J-1)} \frac{1}{N}\sum_{n=1}^{N} \left\Vert 
  {\nabla} f_n(\mathbf{w}^{(t)}) \right\Vert^2
  \\&
  \leq \eta \frac{1}{2} L^2
        \frac{2 \eta^2 \frac{J(J-1)}{2} \sigma_{\mathrm{MM}}^2 }{1-4\eta^2L^2 J (J-1)}
        \\&
        + \eta\frac{1}{2}  L^2\frac{4\eta^2 J^2(J-1)}{1-4\eta^2L^2 J (J-1)}\left(\zeta_1^{\mathrm{AP}} \overline{\Gamma}_{\mathrm{MM}}  \left\Vert {\nabla} f(\mathbf{w}^{(t)})\right\Vert^2 + \zeta_2^{\mathrm{AP}} 
   \right)
    \end{aligned}
    $}
\end{equation}
where the last inequality is obtained based on the gradient diversity (i.e., data heterogeneity) assumption. Replacing the above bound back in~\eqref{pin:2} yields:
\begin{equation}
\hspace{-3mm}
\resizebox{0.89\linewidth}{!}{$
\begin{aligned}
    &\mathbb{E}_t[f(\mathbf{w}^{(t+1)})] \leq  f(\mathbf{w}^{(t)})\\
&-\eta J \frac{1}{2} \left\Vert \nabla f(\mathbf{w}^{(t)})\right\Vert^2+\frac{L}{2}J\eta^2\sigma_{\mathrm{MM}}^2\\
&\eta \frac{1}{2} L^2
        \frac{2 \eta^2 \frac{J(J-1)}{2} \sigma_{\mathrm{MM}}^2 }{1-4\eta^2L^2 J (J-1)}
        \\&
        + \eta \frac{1}{2}L^2\frac{4\eta^2 J^2(J-1)}{1-4\eta^2L^2 J (J-1)}\left(\zeta_1^{\mathrm{AP}} \overline{\Gamma}_{\mathrm{MM}}  \left\Vert {\nabla} f(\mathbf{w}^{(t)})\right\Vert^2 + \zeta_2^{\mathrm{AP}}    \right),
    \end{aligned}
    $}
\end{equation}
which implies that
\begin{equation}\label{pin:4}
\hspace{-3mm}
\resizebox{0.89\linewidth}{!}{$
\begin{aligned}
    &\mathbb{E}_t[f(\mathbf{w}^{(t+1)})] \leq  f(\mathbf{w}^{(t)})\\
&+\eta J \frac{1}{2} \underbrace{\Big( L^2\frac{4\eta^2 J(J-1)}{1-4\eta^2L^2 J (J-1)} \zeta_1^{\mathrm{AP}} \overline{\Gamma}_{\mathrm{MM}} -1\Big)}_{(e)} \left\Vert \nabla f(\mathbf{w}^{(t)})\right\Vert^2\\
&+\eta\frac{1}{2} L^2
        \frac{2 \eta^2 \frac{J(J-1)}{2} \sigma_{\mathrm{MM}}^2 }{1-4\eta^2L^2 J (J-1)}+\frac{L}{2}J\eta^2\sigma_{\mathrm{MM}}^2
        \\&
        + \eta\frac{1}{2} L^2\frac{4\eta^2 J^2(J-1)}{1-4\eta^2L^2 J (J-1)} \zeta_2^{\mathrm{AP}}.
    \end{aligned}
    $}
\end{equation}
We aim to make term $(e)$ negative with  the following condition on the step size:
\begin{equation}\label{pin:3}
    L^2\frac{4\eta^2 J(J-1)}{1-4\eta^2L^2 J (J-1)} \zeta_1^{\mathrm{AP}} \overline{\Gamma}_{\mathrm{MM}} \leq \Omega <1,
\end{equation}
which is satisfied when:
\begin{equation}
\eta \leq \frac{1}{2L} \sqrt{\frac{\Omega }{(\zeta_1^{\mathrm{AP}} \overline{\Gamma}_{\mathrm{MM}}+\Omega )J(J-1)}}.
\end{equation}
Further, under condition~\eqref{pin:3}, we have:
\begin{equation}
     \frac{1}{1-4\eta^2L^2 J (J-1)}  \overline{\Gamma}_{\mathrm{MM}} \leq \frac{\Omega +\zeta_1^{\mathrm{AP}} \overline{\Gamma}_{\mathrm{MM}}}{\zeta_1^{\mathrm{AP}} \overline{\Gamma}_{\mathrm{MM}}} \leq 2.
\end{equation}
Revisiting the bound in~\eqref{pin:4} with the above conditions imply that
\begin{equation}
\begin{aligned}
    &
    \eta J \frac{1}{2} (1-\Omega) \left\Vert \nabla f(\mathbf{w}^{(t)})\right\Vert^2\leq 
     f(\mathbf{w}^{(t)})-\mathbb{E}_t[f(\mathbf{w}^{(t+1)})]\\
&+\eta\frac{1}{2} L^2
        {2 \eta^2 J(J-1) \sigma_{\mathrm{MM}}^2 }+\frac{L}{2}J\eta^2\sigma_{\mathrm{MM}}^2
        \\&
        + \eta\frac{1}{2} L^2 4\eta^2 J^2(J-1)2 \zeta_2^{\mathrm{AP}}.
    \end{aligned}
\end{equation}
Taking the total expectation of both hand sides yields:

\begin{equation}
\begin{aligned}
    &
   \mathbb{E} \Big[ \left\Vert \nabla f(\mathbf{w}^{(t)})\right\Vert^2 \Big]\leq 
     \frac{\mathbb{E}\big[f(\mathbf{w}^{(t)})-f(\mathbf{w}^{(t+1)})\big]}{ \eta J \frac{1}{2} (1-\Omega)}\\
&+\frac{ L^2
        {2 \eta^2 (J-1) \sigma_{\mathrm{MM}}^2 }}{  (1-\Omega)} +\frac{L}{(1-\Omega)}\eta\sigma_{\mathrm{MM}}^2
        \\&
        + \frac{ L^2 4\eta^2 J(J-1)2 \zeta_2^{\mathrm{AP}}}{  (1-\Omega)}.
    \end{aligned}
\end{equation}
Applying $\frac{1}{T}\sum_{t=0}^{T-1}$ to both hand sides and expanding the first term on the right hand side in a telescopic manner (note that three of the right hand side terms are not dependent on $t$, and thus the summation leaves them unchanged), along with using the fact that $f^\star\leq f(w^{T})$, implies:
\begin{equation}
\begin{aligned}
    &
  \frac{1}{T} \sum_{t=0}^{T-1} \mathbb{E} \Big[ \left\Vert \nabla f(\mathbf{w}^{(t)})\right\Vert^2 \Big]\leq 
     \frac{f(\mathbf{w}^{(0)})-f^{\star}}{T \eta J \frac{1}{2} (1-\Omega)}\\
&+\frac{ L^2
        {2 \eta^2 (J-1) \sigma_{\mathrm{MM}}^2 }}{  (1-\Omega)} +\frac{L}{(1-\Omega)}\eta\sigma_{\mathrm{MM}}^2
        \\&
        + \frac{ L^2 4\eta^2 J(J-1)2 \zeta_2^{\mathrm{AP}}}{  (1-\Omega)},
    \end{aligned}
\end{equation}
which proves the main bound in the theorem.

\section{Data Preprocessing}\label{apx:preprocessing}
Here, we explain the preprocessing needed for each dataset to achieve the I/Q modality data which is then used to extract the subsequent modalities and perform the experiments.

\subsubsection{ORBIT Dataset}
\begin{figure}[ht]
    \centering
    \includegraphics[width=\linewidth]{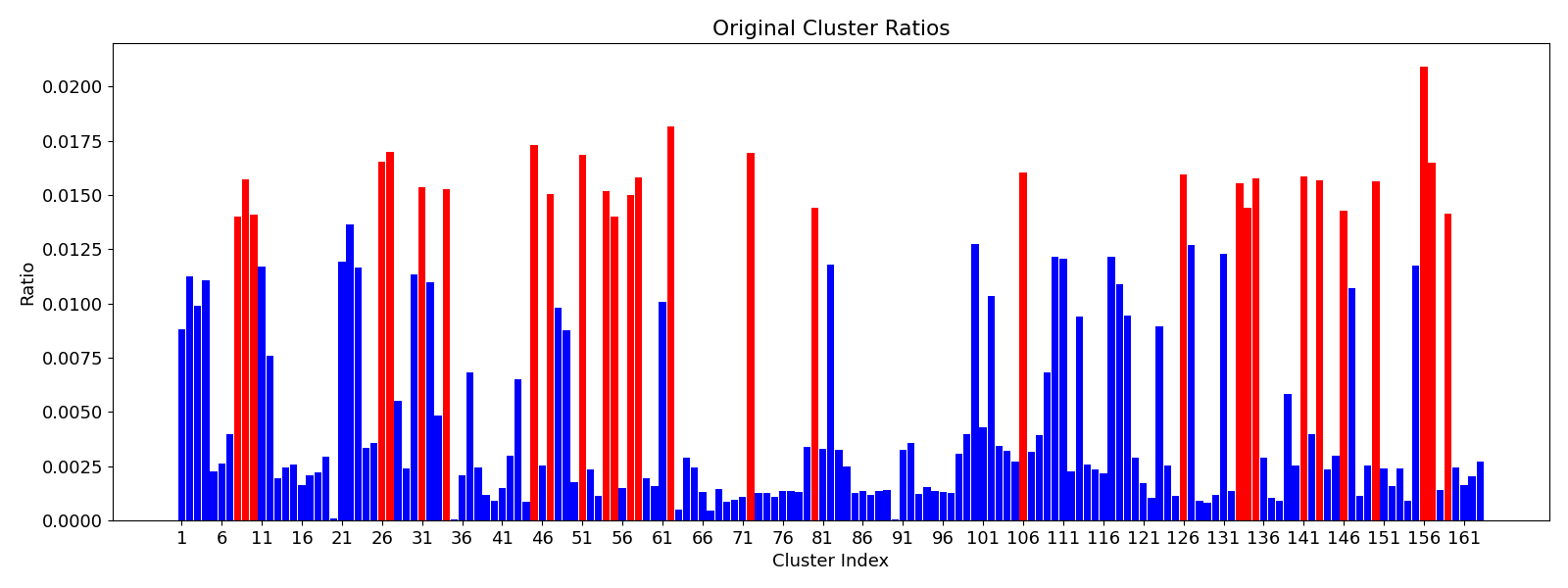}
    \caption{Class distribution for the original ORBIT testbed dataset.}
    \label{fig:orbit-original-dist}
\end{figure}

\begin{figure}[ht]
    \centering
    \includegraphics[width=\linewidth]{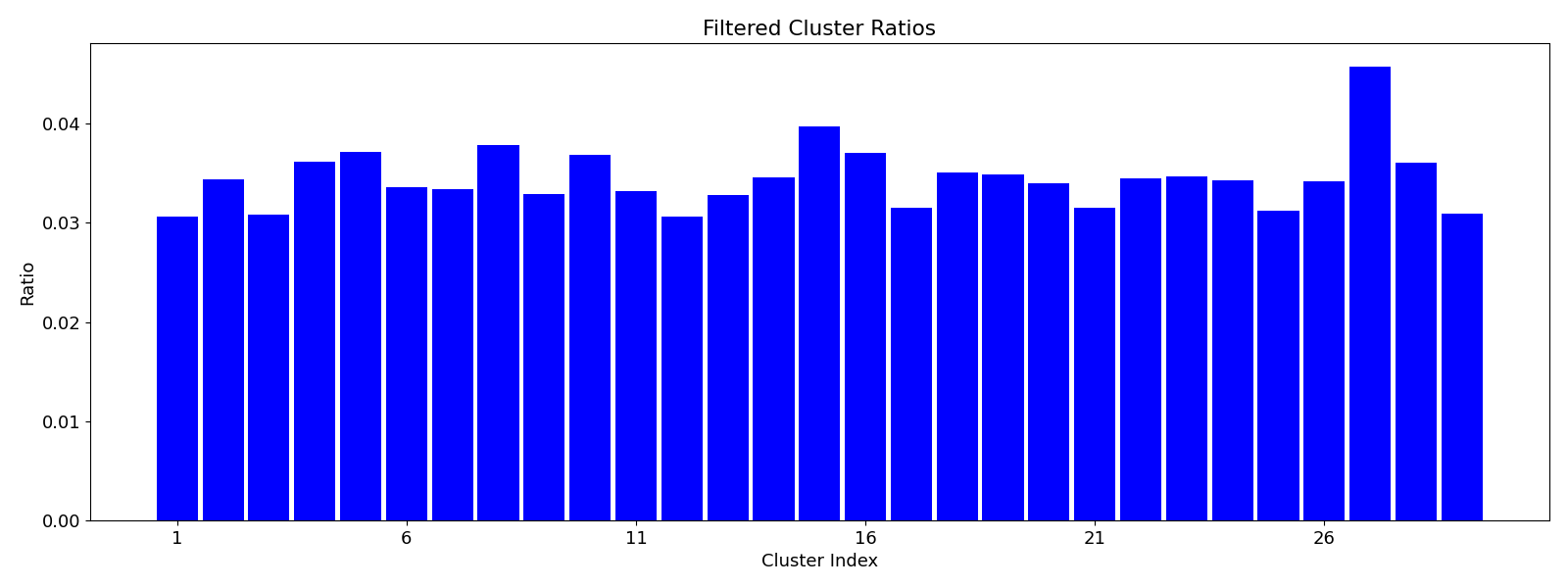}
    \caption{Class distribution for the filtered ORBIT testbed dataset.}
    \label{fig:orbit-filtered-dist}
\end{figure}

In this dataset, there exist 163 classes with their distribution depicted in Fig. \ref{fig:orbit-original-dist}. The observed high imbalance between different data ratios for each class/label can have the following effects on the experiments:
\begin{itemize}
    \item Preventing the realization of a fully IID scenario where the same ratio of each client's dataset is composed of underpopulated classes, due to the lack of enough samples.
    \item Biasing the metrics of the performance of the global model (i.e., accuracy) on the overpopulated classes, undermining the global model performance on the underpopulated classes.
\end{itemize}
To address these issues, we filter the dataset and take the most populated classes (shown in red in Fig. \ref{fig:orbit-original-dist}) to create a balanced subset for our experiments. This filtering will result in the data distribution shown in Fig. \ref{fig:orbit-filtered-dist}.
We then use the filtered dataset with $29$ classes to simulate the clients' local datasets as mentioned in Sec. \ref{datasets}.

\subsubsection{ORACLE Dataset}
In this dataset, the signals received from the transmitters have been recorded under $11$ different distances. For our experiments, we choose the largest distance ($62$ft). Unlike the ORBIT dataset where the preambles of WiFi packets were recorded, the recorded signals in ORACLE are in continuous form. We thus extracted subsamples (i.e., $19,000$ samples from each transmitter) from the continuous waveforms that overlap by $50\%$ with each other, following the work in the original paper introducing the dataset~\cite{sankhe2019oracle}. These subsamples will then be used to extract the frequency modality and amplitude/phase modalities (as explained in Sec.~\ref{mm_rep}) and distributed across clients (explained in Sec.~\ref{exp_setup}).

\begin{IEEEbiography}
[{\includegraphics[width=1in,height=1.25in,clip,keepaspectratio]{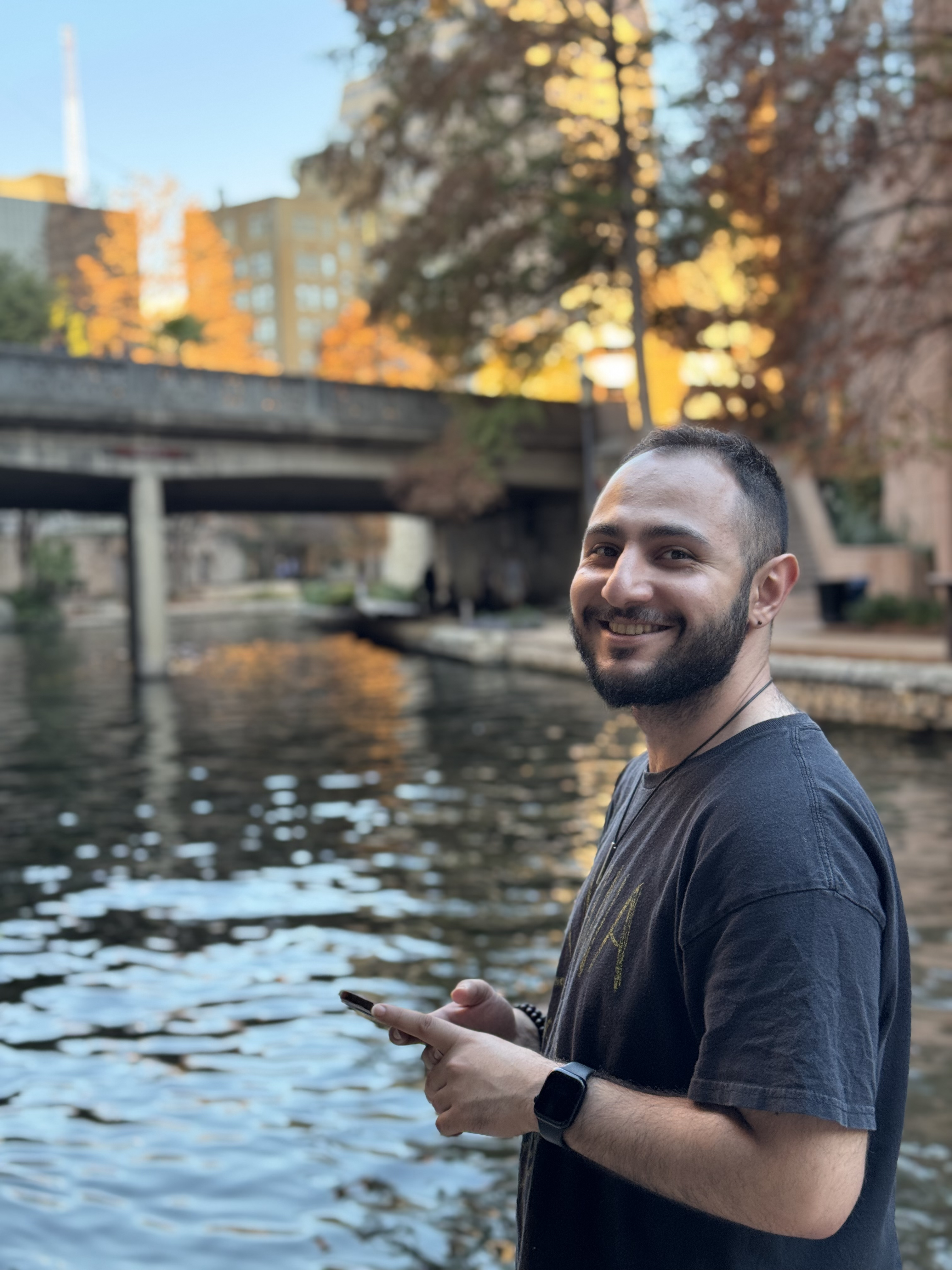}}]{Kasra Borazjani} received his B.Sc. degree in Electrical Engineering from University of Tehran, Tehran, Iran, in 2022. He is currently a Ph.D. candidate in Electrical Engineering at the University at Buffalo--SUNY. His research interests include federated learning (FL), computer vision, and medical image processing, currently focusing on FL in multi-modal and multi-task foundation models.
\end{IEEEbiography}

\begin{IEEEbiography}
[{\includegraphics[width=1in,height=1.25in,clip,keepaspectratio]{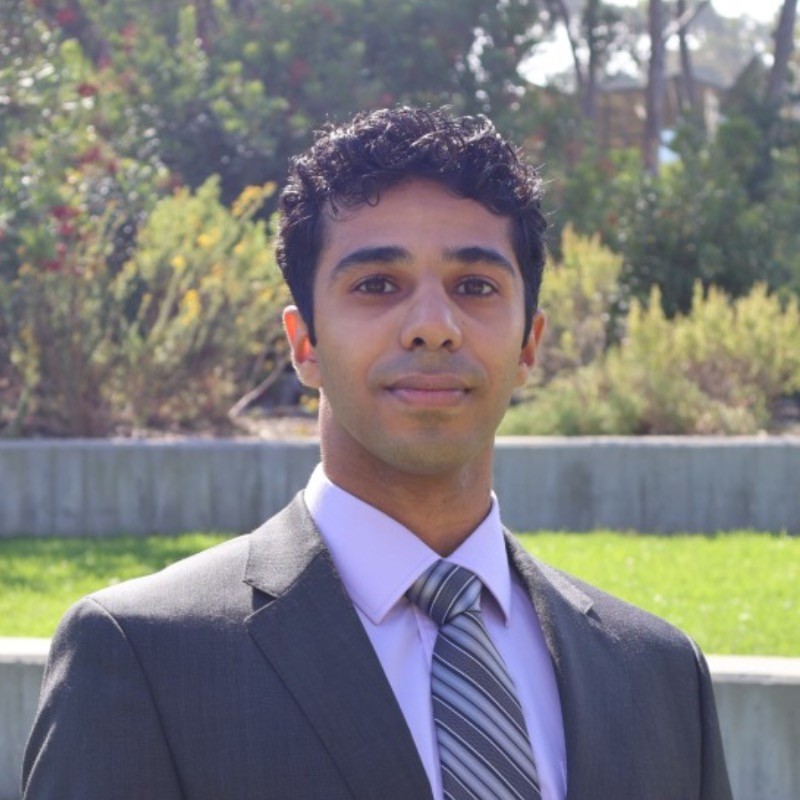}}]{Kiarash Kianfar} is currently pursuing his B.S. in Electrical Engineering at the University of California San Diego. His research interests lie in deep learning, signal processing, and wireless communications.
\end{IEEEbiography}

\begin{IEEEbiography}
[{\includegraphics[width=1in,height=1.25in,clip,keepaspectratio]{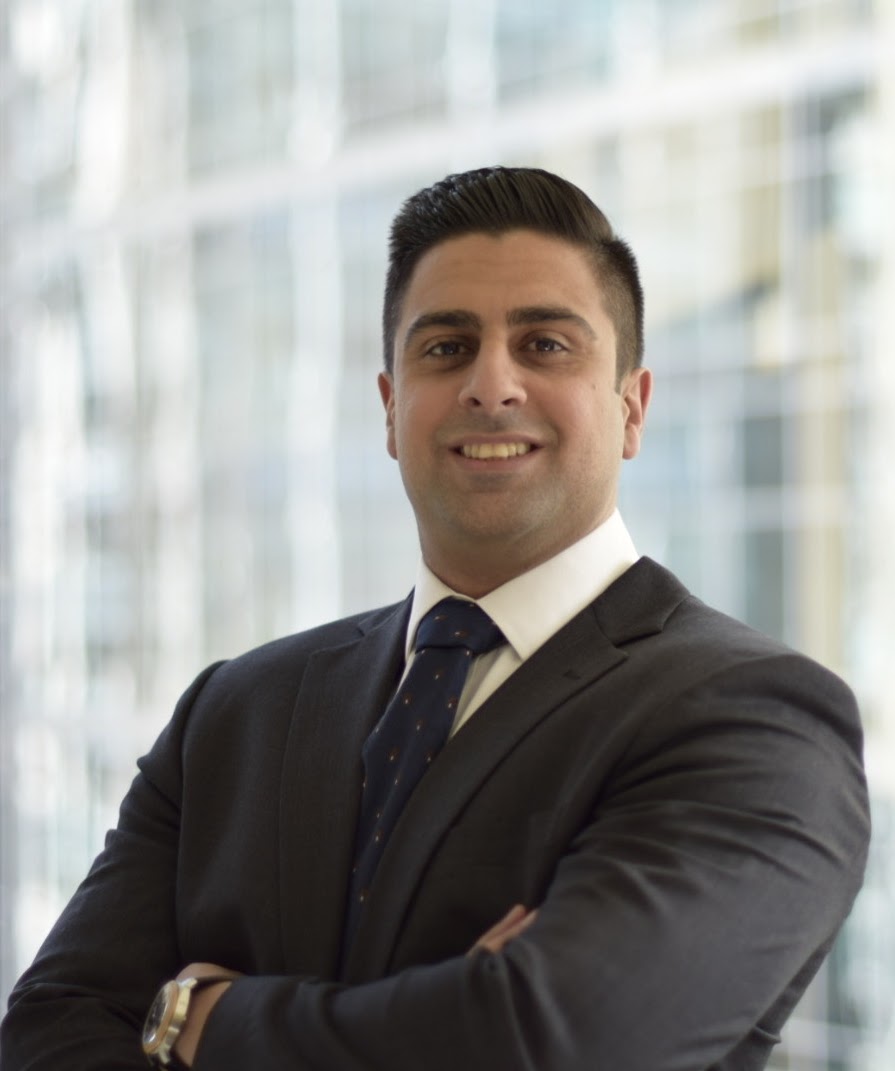}}]{Seyyedali Hosseinalipour}[SM] received the B.S. degree in Electrical Engineering from Amirkabir University of Technology, Tehran, Iran, in 2015 with high honors and top-rank recognition. He then received the M.S. and Ph.D. degrees in electrical engineering from North Carolina State University, Raleigh, NC, USA, in 2017 and 2020, respectively, and was a postdoctoral researcher at Purdue University, West Lafayette, IN, USA, from 2020 to 2022. He was the recipient of the \textit{ECE Doctoral Scholar of the Year Award} (2020) and the \textit{ECE Distinguished Dissertation Award} (2021) at North Carolina State University. Since joining the University at Buffalo (UB)-SUNY, he has received the \textit{National Science Foundation (NSF) CAREER Award} (2026), the \textit{Students' Choice Teaching Excellence Award} (2023), the \textit{School of Engineering and Applied Sciences (SEAS) Early Career Teacher of the Year Award} (2026), and the \textit{2024 IEEE Communications Society William R. Bennett Prize} as the first author of an award-winning paper published in \textit{IEEE/ACM Transactions on Networking}.
He has served as the TPC Co-Chair of workshops and symposiums related to machine learning and edge computing for IEEE INFOCOM, GLOBECOM, ICC, CVPR, ICDCS, SPAWC, WiOpt, and VTC. He also served as a Guest Editor of \textit{IEEE Internet of Things Magazine} for the special issue on \textit{Federated Learning for Industrial Internet of Things} (2023). Since February 2025, he has been serving as an \textit{Associate Editor} for the \textit{IEEE Transactions on Signal and Information Processing over Networks}.
His research interests include the analysis and optimization of modern wireless networks, distributed and federated machine learning, edge/fog computing systems, foundation models for networked intelligence, and network optimization.
\end{IEEEbiography}

\begin{IEEEbiography}
[{\includegraphics[width=1in,height=1.25in,clip,keepaspectratio]{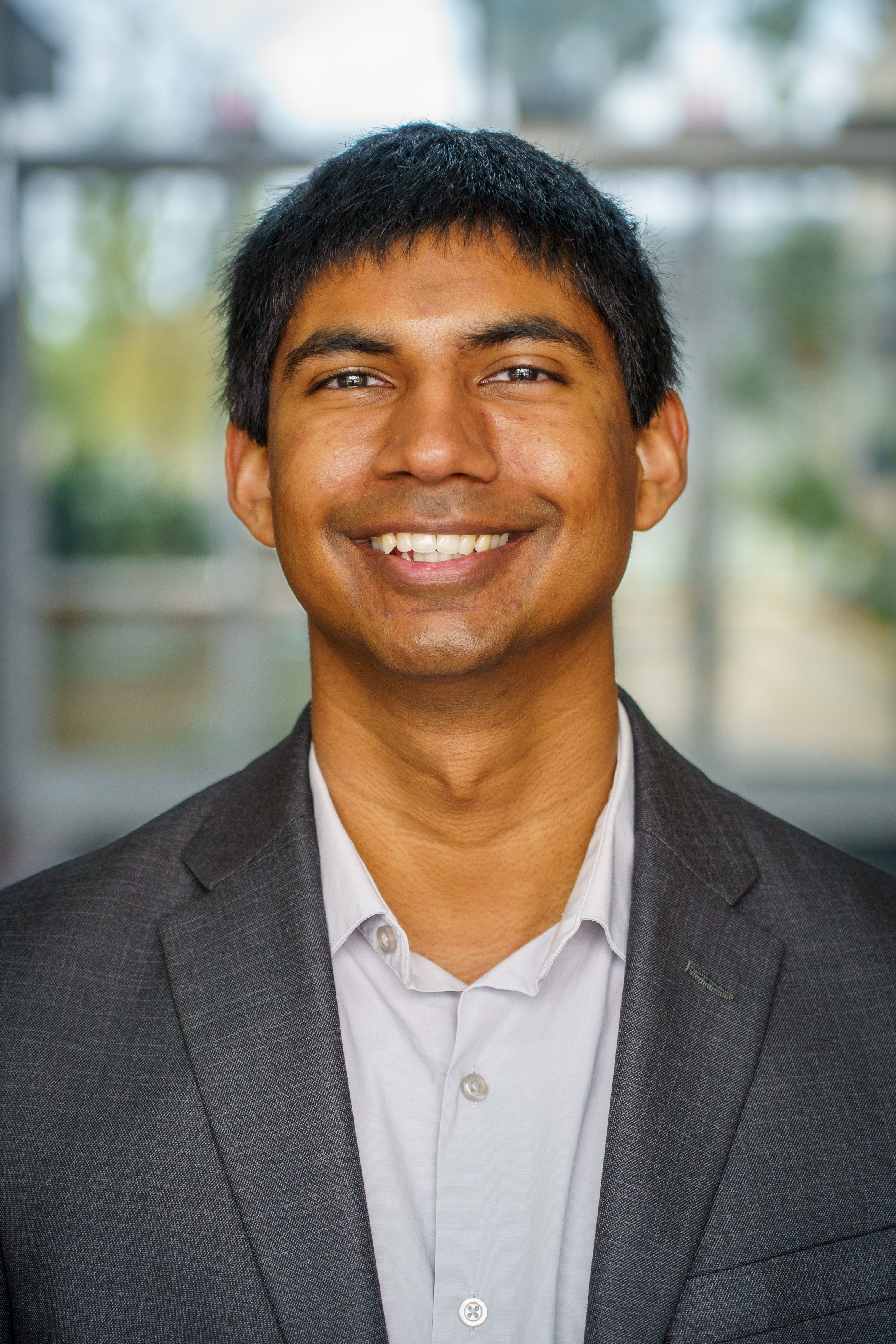}}]{Rajeev Sahay} received the B.S. degree in electrical engineering from The University of Utah, Salt Lake City, UT, USA, in 2018, and the M.S. and Ph.D. degrees in electrical and computer engineering from Purdue University, West Lafayette, IN, USA, in 2021 and 2022, respectively. Currently, he is a faculty member in the Department of Electrical and Computer Engineering at the University of California San Diego (UCSD). He is the recipient of the best undergraduate teaching award in the department of Electrical and Computer Engineering at UCSD, the Purdue Engineering Dean’s Teaching Fellowship and was named an Exemplary Reviewer by the IEEE Wireless Communications Letters. His research interests lie in the intersection of networking and machine learning, especially in their applications to signal processing, wireless communications, and engineering education. 
\end{IEEEbiography}


\end{document}